\documentclass[useAMS,usenatbib]{mnras}

\usepackage{url} 
\usepackage{subfig}
\usepackage{graphicx}
\usepackage{amsmath}

\newcommand\gtsim{\mathrel{\substack{
  \textstyle>\\[-0.2ex]\textstyle\sim}}}

\usepackage{multirow}
\graphicspath{{fig/}} 

\usepackage[utf8]{inputenc}
\usepackage[english]{babel}

\def\aj{AJ}
\def\apj{ApJ}
\def\apjl{ApJ}
\def\apjs{ApJS}
\def\ao{Appl.~Opt.}
\def\apss{Ap\&SS}
\def\aap{A\&A}
\def\aaps{A\&AS}
\def\mnras{MNRAS}
\def\pasp{PASP}
\def\pasj{PASJ}
\def\nat{Nature}

\bibpunct{(}{)}{,}{a}{}{,} 

\title[Blazar AO\,0235+164]{The extreme blazar AO\,0235+164 as seen by extensive ground and space radio observations}

\author[Kutkin et al.]{\parbox{\textwidth}{
A.~M.~Kutkin$^{1}$\thanks{E-mail: kutkin@asc.rssi.ru},
I.~N.~Pashchenko$^{1}$, 
M.~M.~Lisakov$^{1}$, 
P.~A.~Voytsik$^{1}$,
K.~V.~Sokolovsky$^{1,2,3}$, 
Y.~Y.~Kovalev$^{1,4,5}$,
A.~P.~Lobanov$^{5}$,
A.~V.~Ipatov$^{6}$,
M.~F.~Aller$^{7}$,
H.~D.~Aller$^{7}$, 
A.~Lahteenmaki$^{8,9,10}$,
M.~Tornikoski$^{8}$,
L.~I.~Gurvits$^{11,12}$
}
\vspace{0.4cm}\\
\parbox{\textwidth}{
$^{1}$Astro Space Center of Lebedev Physical Institute, Profsoyuznaya Str.
84/32, 117997 Moscow, Russia\\
$^{2}$Sternberg Astronomical Institute, Moscow State University,
Universitetskii~pr. 13, 119992 Moscow, Russia\\
$^{3}$IAASARS, National Observatory of Athens, Vas.~Pavlou \& I.~Metaxa, 15236~Penteli, Greece\\
$^{4}$Moscow Institute of Physics and Technology, Dolgoprudny, Institutsky per., 9, Moscow region, 141700, Russia\\
$^{5}$Max-Planck-Institut f\"ur RadioAstronomie, Auf dem H\"ugel 69,
D-53121 Bonn, Germany\\
$^{6}$Institute of Applied Astronomy of the Russian Academy of Sciences, 191187, Kutuzov embankment, 10, St. Petersburg, Russia
$^{7}$University of Michigan, Astronomy
Department, Ann Arbor, MI 48109-1042 USA\\
$^{8}$Aalto University Mets\"ahovi Radio Observatory,
Mets\"ahovintie 114, 02540 Kylm\"al\"a, Finland\\
$^{9}$Aalto University Dept of Electronics and Nanoengineering,
P.O. Box 15500, 00076 Aalto, Finland\\
$^{10}$Tartu Observatory, Observatooriumi 1, 61602 T\~{o}ravere, Estonia\\
$^{11}$Joint Institute for VLBI ERIC, P.O. Box 2, 7990 AA Dwingeloo, The Netherlands\\%
$^{12}$Department of Astrodynamics and Space Missions, Delft University of Technology, Kluyverweg 1, 2629 HS Delft, The Netherlands\\%
}
}

\begin{document}

\date{Accepted 2018 January 4. Received 2017 December 29; in original form 2017 October 11}

\pagerange{\pageref{firstpage}--\pageref{lastpage}} \pubyear{2018}

\maketitle

\label{firstpage}

\begin{abstract}
Clues to the physical conditions in radio cores of blazars come from measurements of brightness temperatures as well as effects produced by intrinsic opacity. We study the properties of the ultra compact blazar AO\,0235+164 with \textit{RadioAstron} ground-space radio interferometer, multi-frequency VLBA, EVN and single-dish radio observations. We employ visibility modeling and image stacking for deriving structure and kinematics of the source, and use Gaussian process regression to find the relative multi-band time delays of the flares. The multi-frequency core size and time lags support prevailing synchrotron self absorption. The intrinsic brightness temperature of the core derived from ground-based VLBI is close to the equipartition regime value. In the same time, there is evidence for ultra-compact features of the size of less than 10\,$\mu$as in the source, which might be responsible for the extreme apparent brightness temperatures of up to $10^{14}$\,K as measured by \textit{RadioAstron}. In 2007--2016 the VLBI components in the source at 43 GHz are found predominantly in two directions, suggesting a bend of the outflow from southern to northern direction. The apparent opening angle of the jet seen in the stacked image at 43\,GHz is two times wider than that at 15\,GHz, indicating a collimation of the flow within the central 1.5\,mas. 
We estimate the Lorentz factor $\Gamma = 14$, the Doppler factor $\delta=21$, and the viewing angle $\theta = 1.7^\circ$ of the apparent jet base, derive the gradients of magnetic field strength and electron density in the outflow, and the distance between jet apex and the core at each frequency.
\end{abstract}


\section{Introduction}
\label{sec:intro}

Blazars are jetted active galactic nuclei (AGN) viewed at a small angle to the line of sight. They appear in very long baseline interferometry (VLBI) images as an unresolved or barely resolved bright core often accompanied by a one-sided jet. Radio emission is produced via the synchrotron mechanism and is boosted by relativistic effects.
The core has a flat or inverted spectrum at cm/mm wavelengths and is usually associated with emission from a surface of unit optical depth \citep{1979ApJ...232...34B}. The alternative interpretation of the core at mm and sub-mm wavelengths involve shock models~\citep{2006AIPC..856....1M,2008ASPC..386..437M}.
Strong support for the former interpretation comes from the shift between the apparent core positions measured at different frequencies, first detected by \cite{1984ApJ...276...56M}. 
In the framework of this model, the measured core shift can be used to probe jet physics in the core region including the magnetic field strength and its distribution along the outflow, the linear scale of the jet, the jet kinetic energy, etc. \citep{1998A&A...330...79L, 2005ApJ...619...73H}. The core-shift effect should also be taken into account while contracting and aligning precise celestial reference frames \citep[e.g.,][]{2017A&A...598L...1K,2017MNRAS.467L..71P,2017MNRAS.471.3775P}

The core shift can be measured directly if a source has a prominent jet, so that one can use its optically thin features as a reference~\citep[e.g., ][] {2008A&A...483..759K}. There is also an astrometric method to measure the core shift using nearby sources for reference \citep[e.g.,][]{2011Natur.477..185H}. 
For a compact source the core shift can be estimated indirectly. Opacity is also responsible for changing the angular size of the apparent core measured at different frequencies. Therefore the dependence of core size on frequency can be linked to the core shift with assumptions about geometry of a jet. Moreover, these measurements allow one to distinguish between intrinsic absorption and scattering in the interstellar medium \citep[e.g.,][]{2015MNRAS.452.4274P}. Finally, the time delays between single-dish light curves at radio bands might also reflect the offset between the core at these frequencies and the outflow velocity \citep{2011MNRAS.415.1631K,2014MNRAS.437.3396K}.

The radio source AO\,0235$+$164 (hereafter 0235+164; 02:38:38.930107 +16:36:59.27460\footnote{\url{http://astrogeo.org/vlbi/solutions/rfc_2016c/}}, J2000) was classified as a BL\,Lac-type object by \cite{1975ApJ...201..275S} on the basis of its variability and optical spectrum, which appeared featureless at low spectral resolution. \citet{2010A&A...518A..10V} classify the object as QSO based on its absolute magnitude. 
\cite{1987ApJ...318..577C} measured the object's redshift detecting Mg\,II, [Ne\,V] and [O\,II] lines at $z=0.94$. They also found intervening absorption and emission features at lower redshifts. 
The blazar resides in a field of many faint foreground galaxies \citep{1996AJ....112.2533B,1996A&A...314..754N} with $z\sim0.5$, and is often considered to be affected by gravitational lensing \citep{1988A&A...198L..13S, 1993ApJ...415..101A, 2000AJ....120...41W}. 

As seen by VLBI, the source is partially resolved at most radio bands, however its extended structure is very unstable. \citet{2000MNRAS.319.1056G} report no detectable
milliarcsecond-scale structure, while in other works, some hints of jet-like morphology are reported (see, e.g., \citealt{1984ApJ...284...60J, 1995A&AS..114..197A, 1996A&A...307...15C}, and references therein). 
\cite{2001ApJS..134..181J} detected the jet north-west of the core at 43\,GHz and measured the apparent superluminal speed of two components. Later 43\,GHz VLBA monitoring revealed the temporary appearance of a component located south-southeast of the core (\citealt{2011ApJ...735L..10A}, this paper). 

The blazar 0235+164 is one of the few brightest sources detected by VSOP and \textit{RadioAstron} space-VLBI missions. Based on VSOP measurements, its brightness temperature is reported to reach $T_\mathrm{b} \approx 10^{13.8}$\,K~\citep{2000PASJ...52..975F}, which challenges the inverse Compton limit even after being corrected for boosting with extremely high Doppler factors.

The source demonstrates violent variability across the electromagnetic spectrum on time-scales from hours to years (e.g., \citealt{2012ApJ...751..159A} and references in their introduction). The observed cm-wavelength flux density of 0235+164 increases up to 6-7 times during the flaring states. 
The short-term variability at lower frequencies might be produced by inter-stellar scintillation of newborn ultra-compact VLBI components with a size about 10 $\umu$as~\citep{2006ApJS..165..439R}. These blobs, in turn, might be responsible for the extremely high brightness temperatures in the source.

In this paper we analyze both VLBI, including Space-VLBI \textit{RadioAstron} data, and single-dish total flux radio observations of 0235$+$164 to measure multi-frequency time lags and core size, derive the brightness temperature and Doppler factor, study the jet structure and estimate its physical and geometrical parameters. 

We adopt the standard $\Lambda$CDM cosmology model with 
$H_0 = 71$\,km\,s$^{-1}$\,Mpc$^{-1}$,
$\Omega_\mathrm{m} = 0.27$, and $\Omega_{\Lambda} = 0.73$
\citep{2009ApJS..180..330K}, which corresponds to
a luminosity distance $D_L = 6141$\,Mpc,
an angular size distance $D_A = 1632$\,Mpc,
and a linear scale of 7.91~pc\,mas$^{-1}$ at the source redshift.
We use positively-defined spectral index $\alpha=d \ln S/d \ln \nu$.

\begin{figure*}
	\includegraphics[width=5.3cm]{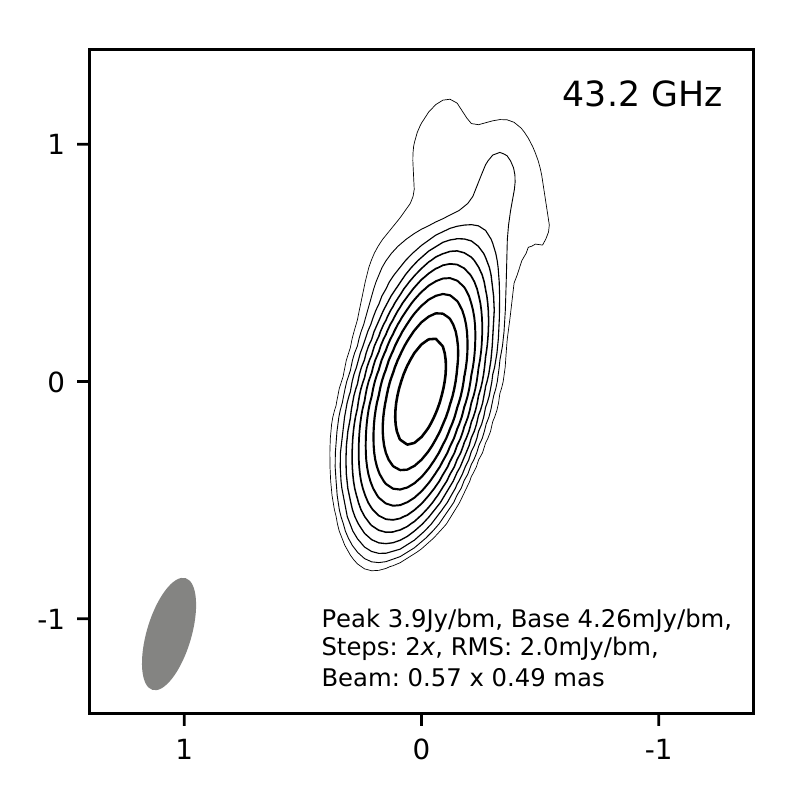} 
	\includegraphics[width=5.3cm]{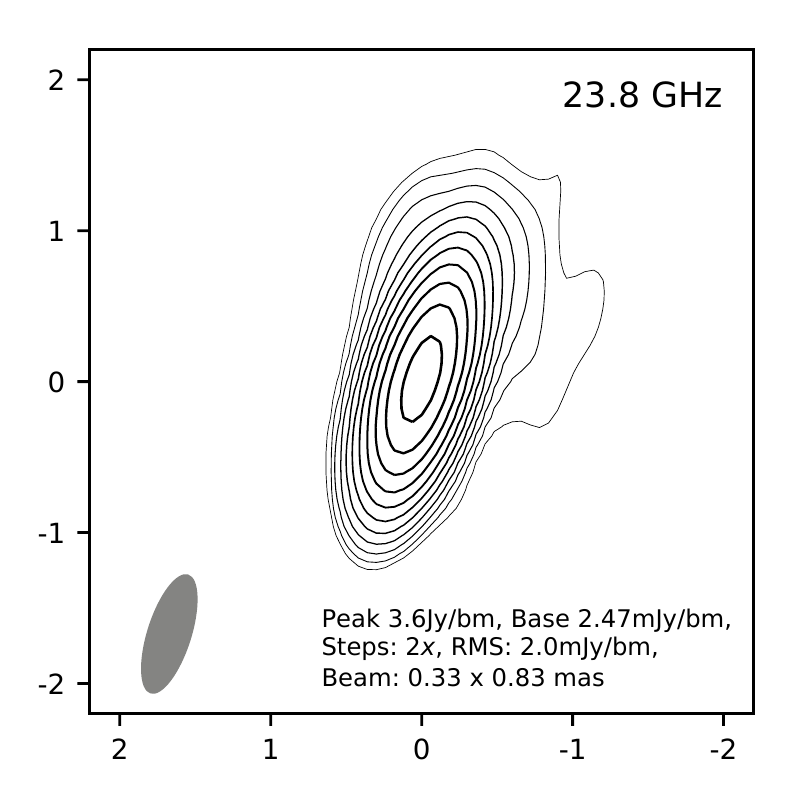} 
	\includegraphics[width=5.3cm]{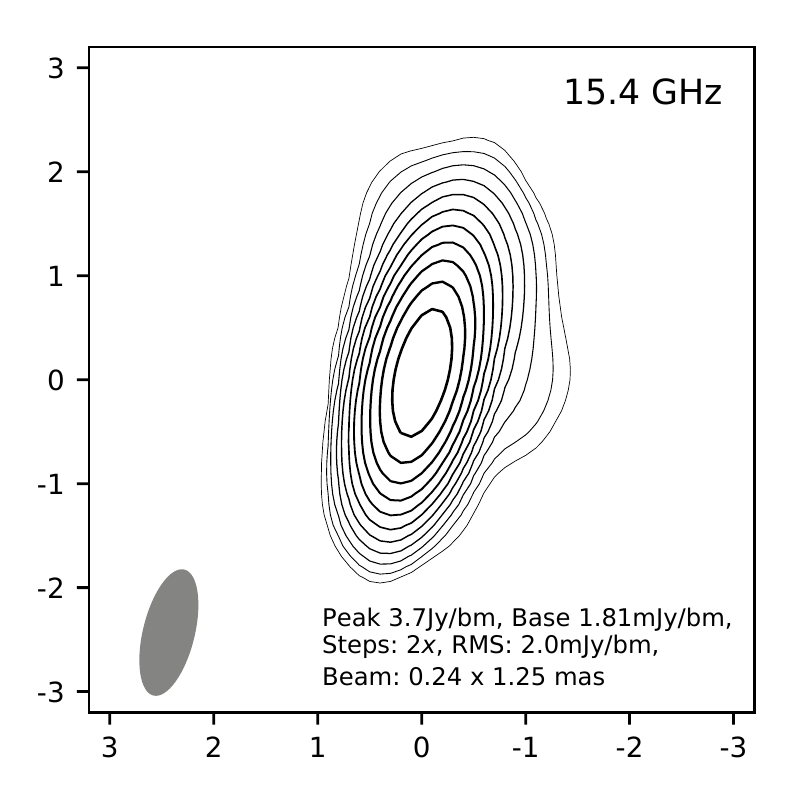} 
	\includegraphics[width=5.4cm]{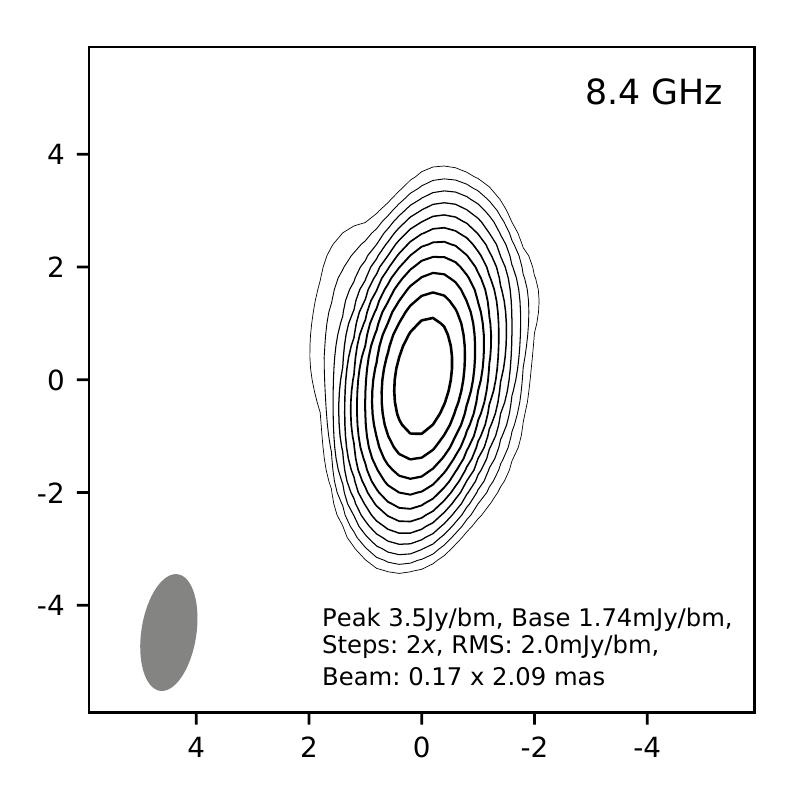}
	\includegraphics[width=5.4cm]{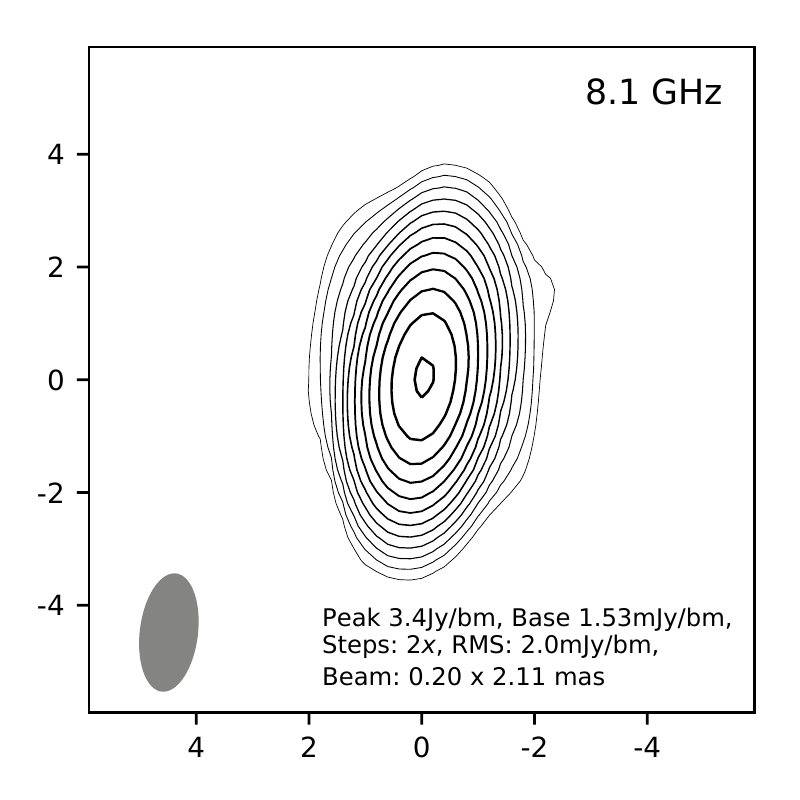} 
	\includegraphics[width=5.4cm]{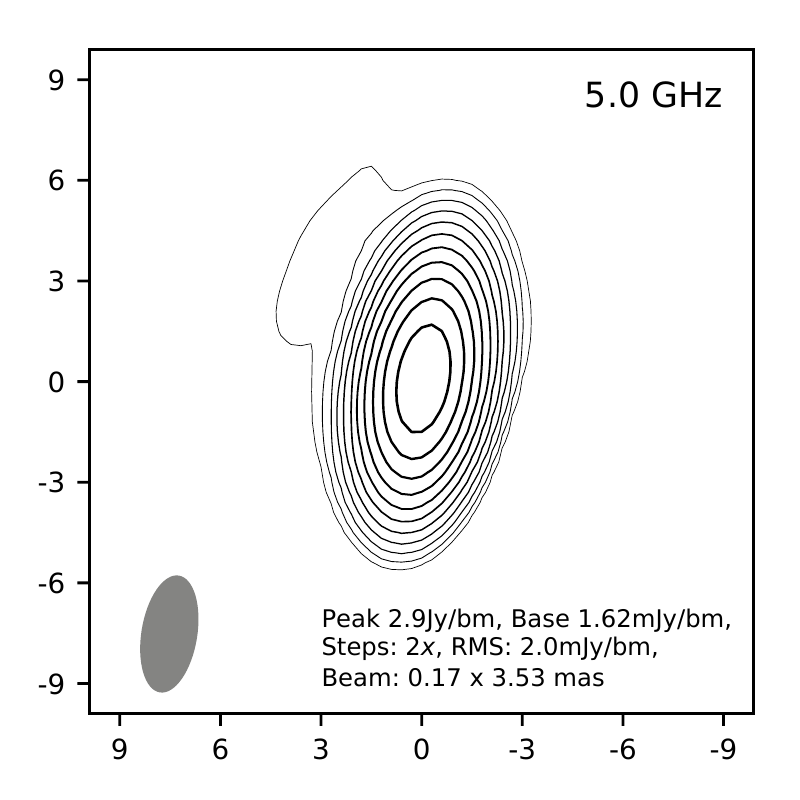} 
	\includegraphics[width=5.4cm]{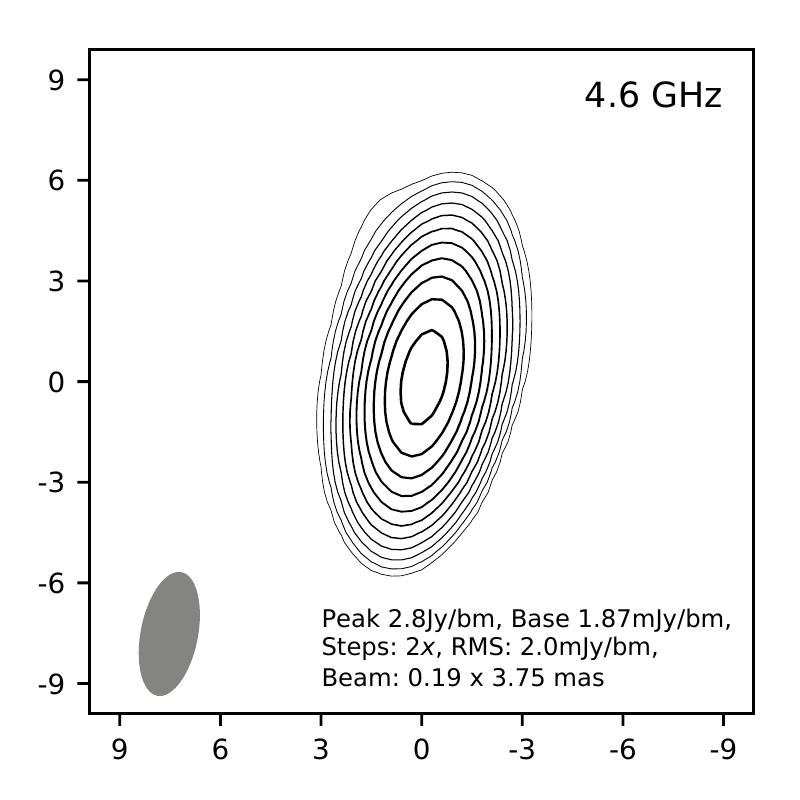}
	\includegraphics[width=5.4cm]{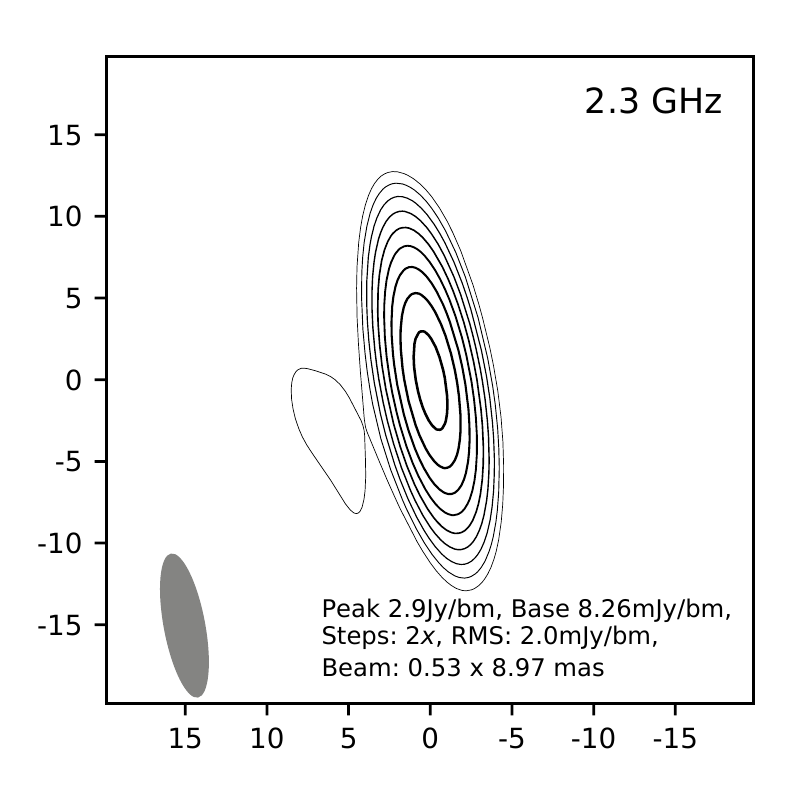} 
	\includegraphics[width=5.4cm]{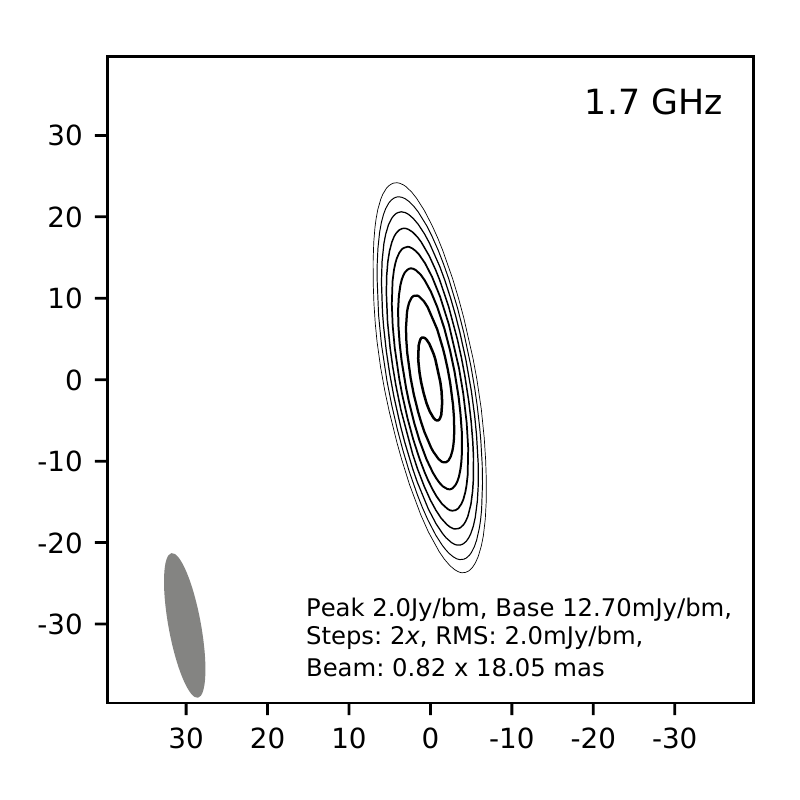} 
	\caption{0235$+$164 clean maps in RA-Dec axes. The data at 4.6--43.2\,GHz are by VLBA, at 1.7/2.3\,GHz -- by EVN. The beam size at half maximum is plotted in the lower left corner.}
	\label{fig:cleanmaps}		
\end{figure*}

\begin{figure*}
	\includegraphics[width=\textwidth,trim=0cm 1cm 0cm 0cm]{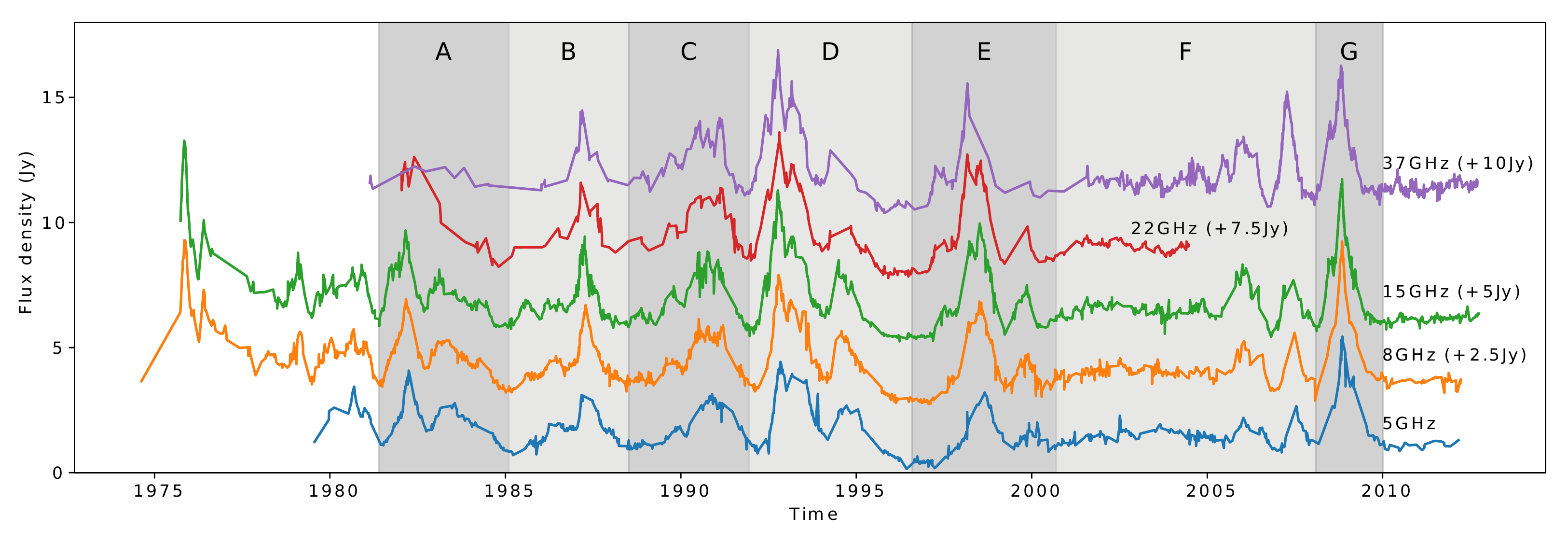}
	\caption{The single-dish total flux density light curves of AO\,0235+164 at 4.8\,GHz to 37\,GHz shifted along y-axis for better visualization. Shaded areas denote the epochs of interest, see Section~\ref{sec:gpr} for details.}
	\label{fig:lightcurves}
\end{figure*}


\section{Observational data}
\label{sec:obsdata}

\subsection{Ground-based VLBI observations}

The blazar 0235$+$164 was observed on 2~September 2008 with the Very Long Baseline Array\footnote{\url{https://science.nrao.edu/facilities/vlba}} (VLBA; \citealt{1994IAUS..158..117N}), providing baselines of up to 8600\,km. The VLBA's frequency agility was used to perform imaging simultaneously at seven frequencies
($4.6$, $5.0$, $8.1$, $8.4$, $15.4$, $23.8$, and $43.2$\,GHz). Eight 8\,MHz-wide frequency channels (IFs) were recorded at 128\,Mbps with 1-bit sampling. The 5\,GHz and 8\,GHz data were split into two sub-bands (four IFs each) to provide independent measurements at close frequencies.
The observation included 13 on-source scans,
each 3-7~minutes long depending on frequency, spread over 8~hours.
The data were correlated at the NRAO Array Operation Center in Socorro, NM.
This observation was conducted in
the framework of our survey of parsec-scale radio spectra of twenty
$\gamma$-ray bright blazars \citep{2010arXiv1006.3084S,2010arXiv1001.2591S}.

Another set of multi-frequency observations was performed by the European VLBI Network at 2.3\,GHz and 8.4\,GHz on 19~October 2008, 5\,GHz on 22~October 2008 and 1.7\,GHz on 29~October~2008. Each band included 8 frequency channels of 8~MHz width each. The 5/1.7~GHz bands were recorded in right and left circular polarizations, while 8.4/2.3\,GHz bands where recorded in right circular polarization with a full bitrate of 512~Mbit/s. The correlation was performed at the Joint Institute for VLBI ERIC~\citep{2015ExA....39..259K}. 

The a priori amplitude calibration, phase calibration with the phase-cal signal injected during observations, fringe fitting (performed separately for each IF), and bandpass correction were applied in \texttt{AIPS} \citep{1990apaa.conf..125G}.
The hybrid imaging \citep{1995ASPC...82..247W} including iterations of image deconvolution using the \texttt{CLEAN} algorithm \citep{1974A&AS...15..417H} followed by amplitude and phase self-calibration were performed in \texttt{Difmap} \citep{1997ASPC..125...77S}.
We applied a special procedure involving preliminary imaging used to determine residual antenna gain corrections that are persistent in time and appear for all the observed sources (a similar procedure was utilized by \citealt{2011A&A...532A..38S}). The resulting amplitude calibration accuracy is expected to be $\sim 5$\,\% in the $4.6$\,GHz to $15.4$\,GHz range and
$\sim 10$\,\% at $23.8$\,GHz and $43.2$\,GHz. Details of the calibration and analysis techniques are described by \cite{2011PhDT.........6S}.

The clean maps of 0235+164 at different frequencies are shown in Figure~\ref{fig:cleanmaps}. Both multi-frequency experiments where made during a prominent radio flare in the source. 

We have also re-imaged and analyzed the calibrated VLBA $uv$-data at 43~GHz by the Boston University blazar group\footnote{\url{http://www.bu.edu/blazars/VLBA_GLAST/0235.html}} covering 2007--2016 (100 observational epochs). 
The imaging and model fitting procedures were performed in the same manner as for our multi-frequency data.

\subsection{Single-dish total flux density observations}

The single-dish flux density monitoring observations (Figure~\ref{fig:lightcurves}) of 0235$+$164 were obtained
with the 26m radio telescope of the University of Michigan Radio Observatory\footnote{\url{http://lsa.umich.edu/astro}} (UMRAO) at 4.8\,GHz, 8.0\,GHz and 14.5\,GHz, with the 40m telescope of Owens Valley Radio Observatory\footnote{\url{http://www.astro.caltech.edu/ovroblazars/}} (OVRO) at 15\,GHz~\citep{2011ApJS..194...29R}, and the 14m telescope of Mets\"ahovi Radio Observatory\footnote{\url{http://metsahovi.aalto.fi/en/}} at 22\,GHz and 37\,GHz. For the subsequent analysis we merge the complementary 15\,GHz data by OVRO and UMRAO.


\subsection{RadioAstron space-VLBI observations}
\label{sec:ra_obs}

The blazar 0235+164 is a target of the \textit{RadioAstron} AGN Survey (Kovalev et al., in prep). It is monitored regularly at 1.6\,GHz, 4.8\,GHz and 22.2\,GHz by the space radio telescope antenna along with ground-based telescopes in interferometric mode. A typical observational set covers 40--60 minutes and is performed either in a single band with two circular polarizations or in two bands with one circular polarization of opposite sense per band. Correlation was performed at the Astro Space Center using a software correlator~\citep{2013ARep...57..153K,2017JAI.....650004L}. Post-correlation handling (fringe fitting, calibration) was performed using \texttt{PIMA}\footnote{\url{http://astrogeo.org/pima}} software~\citep{2011AJ....142...35P}.
We analyzed 12 observational epochs from Dec 2012 through Jan 2016, as summarized in Table~\ref{tab:ra}.


\section{VLBI analysis}
\label{sec:vlbi}

We modeled the VLBI structure of 0235+164 at each frequency in the $uv$-plane using \texttt{Difmap}. 
The $K$-fold cross-validation~\citep{esl} was used to select the best model among a point source, elliptical and circular Gaussian shape of the core (the other components, if any, were modeled with circular Gaussian profiles).
This method is aimed to estimate prediction performance of a model on a new data based on the data at hand.
It employs the data splitting into $K$ non-overlapping sub-samples. Then the following steps are done: one sub-sample is excluded from the data, the model is fitted to the rest $K-1$ sub-samples, and its prediction performance (score) for the excluded subset is evaluated (e.g. by calculating residuals RMS). This procedure is repeated $K$ times with changing the excluded subset. The averaged score of the obtained $K$ values characterizes how the model can generalize (i.e. describes an arbitrary new data), and is used for comparison of different models.
We apply this method to the calibrated $uv$-data and the models constructed in \texttt{Difmap}. To ensure that all the baselines are represented evenly during the procedure we performed the splitting at each one separately. 
We also varied $K$ in the range $3-10$ for the sake of robustness. The elliptical Gaussian model was found to better describe the core. We note, that the extended structure, if any, is weak and does not affect core parameters significantly. The errors in the model parameters were estimated in the image plane following~\cite{1999ASPC..180..301F}.  

Additionally we compared the size of the fitted core with the resolution limit suggested by \cite{2005AJ....130.2473K} for a Gaussian brightness distribution template. The core is resolved at all frequencies except for 1.7~GHz and at almost all epochs of the 7~mm long-term monitoring data. If the core was unresolved, we used the corresponding resolution limit for its major and minor axes size. Models for the multi-frequency experiments are summarized in Table~\ref{tab:difmap_models}.

\begin{table}
	\renewcommand{\arraystretch}{1.5}
	{\scriptsize
	\caption{Source models for the two multi-frequency experiments. Columns: (1)-- frequency, (2) -- component ID, (3--8) -- standard \texttt{Difmap} format: Flux density, Distance, Position angle, Major axis, Axial ratio, Major axis orientation angle.}\label{tab:difmap_models}  
	\begin{tabular}{|c|c|c|c|c|c|c|c|}
		\hline 
		Freq. & ID & $S$  & $R$   & $\theta_c$ 	& $b_{maj}$ & $e$	& $\Phi$  \\ 
		(GHz) &	   & (Jy) & (mas) & (deg) 		& (mas) 	&  		& (deg) \\
		(1)   & (2)& (3)  &   (4) & (5)   		& (6)   	& (7) 	& (8) \\
		\hline 
		43.2 	& C0 & $3.498$ & $0.027$ & $ 171$ & $0.058$ & $0.56$ & $7.7$ \\ 
				& C1 &  $0.636$ & $0.146$ & $-178$ & $0.064$&& \\
				& C2 & $0.051$ & $0.149$ & $ 113$ & $<0.001$&& \\
				& C3 & $0.062$ & $0.613$ & $ -18$ & $0.837$&& \\
		23.8 	& C0 & $3.714$ & $0.015$ & $   2$ & $0.115$ & $0.43$ & $24.0$ \\ 
				& C1 & $0.046$ & $0.204$ & $  88$ & $0.174$&& \\
				& C2 & $0.110$ & $0.634$ & $ -22$ & $0.714$&& \\
		15.37 	& C0 & $3.714$ & $0.059$ & $  -2$ & $0.125$ & $0.50$ & $28.0$ \\
				& C1 & $0.163$ & $0.725$ & $ -18$ & $0.821$&& \\
		8.43 	& C0 & $3.378$ & $0.038$ & $ -28$ & $0.197$ & $0.56$ & $-13.6$ \\
				& C1 & $0.258$ & $0.686$ & $ -23$ & $0.802$&& \\
		8.11 	& C0 & $3.288$ & $0.035$ & $ -51$ & $0.196$ & $0.45$ & $-26.3$ \\
				& C1 & $0.247$ & $0.729$ & $ -23$ & $0.772$&& \\
		5.0 	& C0 & $2.758$ & $0.057$ & $ -42$ & $0.328$ & $0.57$ & $-10.9$ \\
				& C1 & $0.288$ & $0.669$ & $ -22$ & $1.127$&& \\
		4.61 	& C0 & $2.642$ & $0.097$ & $ -39$ & $0.346$ & $0.51$ & $-16.8$ \\
				& C1 & $0.297$ & $0.678$ & $ -22$ & $1.189$&& \\
		\hline		
		 8.39 	& C0 & $6.457$ & $0.049$ & $-173$ & $0.197$ & $0.56$ & $5.0$ \\
				& C1 & $0.374$ & $0.447$ & $ -21$ & $0.655$&& \\
				& C2 & $0.015$ & $1.550$ & $  79$ & $1.001$&& \\
		4.97	& C0 & $4.897$ & $0.032$ & $ 143$ & $0.290$ & $0.46$ & $0.7$ \\
				& C1 & $0.414$ & $0.588$ & $ -16$ & $1.003$&& \\
		2.27 	& C0 & $2.987$ & $0.057$ & $-163$ & $0.772$ & $0.73$ & $-11.0$ \\
				& C1 & $0.015$ & $2.864$ & $ -48$ & $<0.001$&& \\
				& C2 & $0.037$ & $6.164$ & $ 112$ & $5.684$&& \\
		1.66 	& C0 & $2.069$ & $0.093$ & $  19$ & $<1.590$ & $0.60$ & $2.1$ \\
				& C1 & $0.010$ & $4.315$ & $  61$ & $<0.001$&& \\
				& C2 & $0.010$ & $4.902$ & $ -24$ & $0.009$&& \\
		
		\hline	
	\end{tabular}
}

\end{table}


\section{Light curves analysis}
\label{sec:gpr}

The variability time delays between light curves can be found using several methods based either on data modeling or cross-correlation techniques. The former implies some prior knowledge about the data \citep[e.g., Gaussian shapes of the flares,][]{2007MNRAS.381..797P, 2011MNRAS.415.1631K} or two-sided exponential profiles \citep{1999ApJS..120...95V, 2009A&A...494..527H}. Moreover, one has to manually distinguish separate flares profiles, which is the subject of errors due to the Human factor. The latter methods are based on calculation of mutual cross-correlation function (CCF) of the light curves and do not involve additional information about their structure (e.g., standard correlation function with interpolation (ICCF) or discrete correlation function \citep[DCF,][]{1988ApJ...333..646E}. However these methods require a sufficient amount of data to perform well and also require some parameters choice that may affect the results, like the bin width used for interpolation and DCF evaluation (see \citealt{1994PASP..106..879W} for ICCF/DCF comparison and references). 

Another approach, which also does not require any prior knowledge about the data is based on the Gaussian process regression \citep[GPR,][]{Rasmussen:GPM} -- a special case of Bayesian non-parametric models. It has been recently applied to the light curves of the blazar PKS\,1502+106 by \citet{2016A&A...590A..48K}. 

Gaussian process (GP) is a probability distribution over functions. It is characterized by its mean value and covariance matrix. The former can be set to zero by shifting the data, the latter is expressed through the covariance function (\textit{kernel}). The kernel depends on \textit{hyperparameters}, which are inferred for a given data set by maximizing the marginal likelihood function (\textit{training} the GP). 

Any GP realization can be considered as a function (in case of time series this is the function of time). The often-used GP kernel for time series is the ``squared exponential'' (SE):

\begin{equation}
    C_{SE}(t_i,t_j) = A^2\exp\left(\frac{-(t_i - t_j)^2}{2l^2}\right),
    \label{eq:kse}
\end{equation}
where C -- is the covariance between the function values at arbitrary times $t_i$ and $t_j$, $l$ and $A$ are the hyperparameters that correspond to the  characteristic time-scale and amplitude of function variations. 

In the case of longer light curves containing several or multi-peaked flares the single SE kernel~(\ref{eq:kse}) might not perform well due to the presence of more than one characteristic scale in the source variability. This is the case for blazars, which are known to vary on a range of time-scales and usually possess ``red'' power spectra
(e.g., \citealt{2014MNRAS.445..437M}, and references therein). 
To overcome this problem one may use a sum of several SE-kernels. Such a  weighted sum is represented by the rational quadratic (RQ) kernel \cite[Chapter 4]{Rasmussen:GPM}:

\begin{equation}
    C_{RQ}(t_i,t_j) = A^2\left(1 + \frac{(t_i - t_j)^2}{(2\varepsilon l^2)}\right)^{-\varepsilon}
    \label{eq:krq}
\end{equation}

The relative weighting of different time scales variations is specified by the hyper parameter $\varepsilon \in (0, +\infty)$. In the limit of $\varepsilon \rightarrow +\infty$ the RQ kernel becomes an SE kernel. 

We use a kernel represented by a sum of the RQ-kernel and the White kernel characterizing the noise:
\begin{equation}
    C(t_i, t_j) = C_{RQ}(t_i, t_j) + \delta_{ij} (\sigma_{obs}^{2} + \sigma_n^{2})
	\label{eq:kernel}
\end{equation}
with the hyperparameters $A, l, \varepsilon$ of the RQ-kernel. $\sigma_{obs}$ are the measurements uncertainties of the data, and the additional $\sigma_n$ is used to describe the possible unaccounted noise (\textit{jitter}). $\delta_{ij}$ is the Kronecker delta.

\begin{figure*}
	\includegraphics[width=\textwidth,trim=0cm 0.7cm 0cm 0cm]{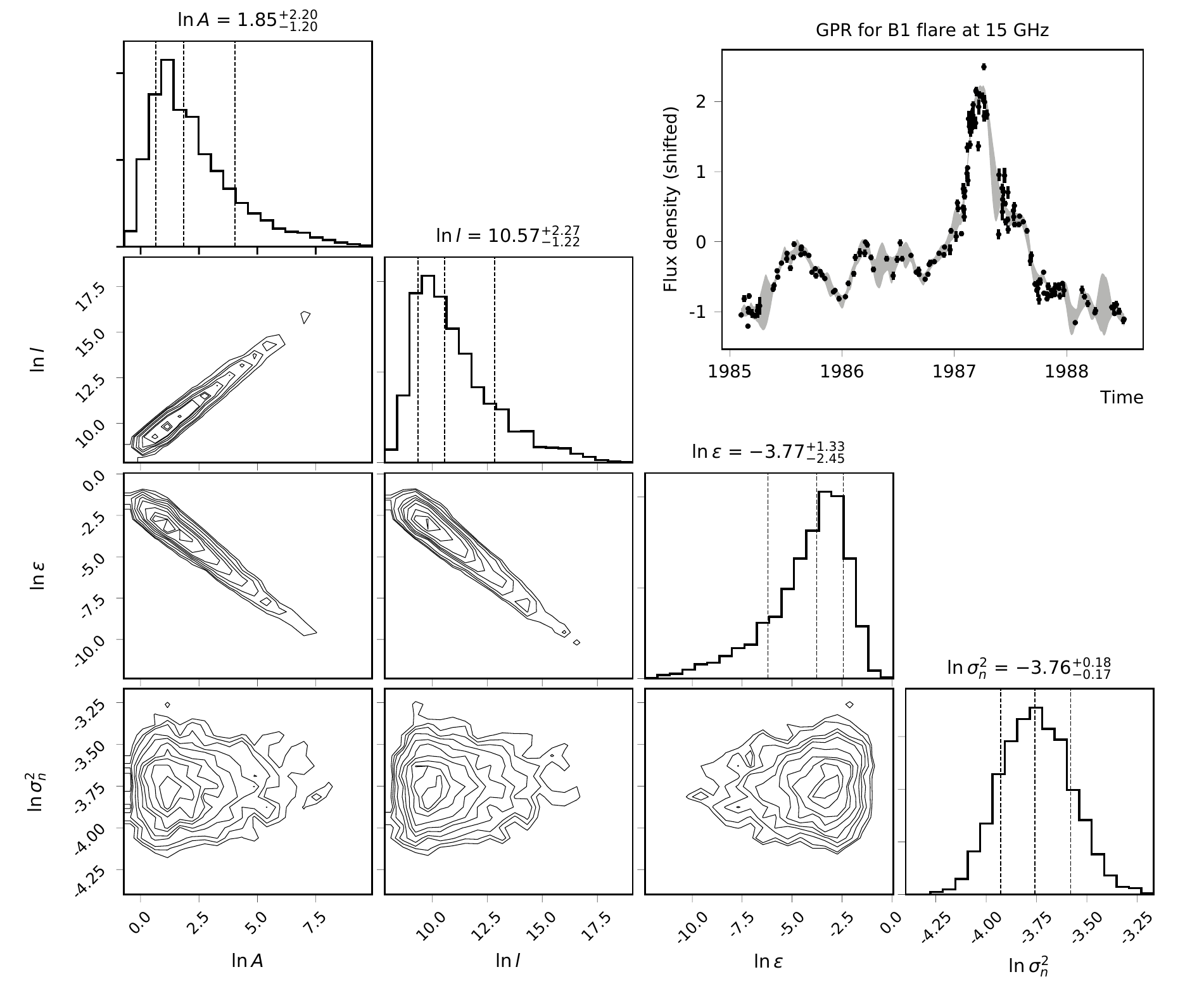}
	\caption{Application of GPR to the 15 GHz B-flare: the posterior distributions of GP hyperparameters (dashed lines indicate 16, 50 and 84 percentiles). The light curve with mean flux density subtracted is shown in the upper right corner, shaded area  denotes $\pm 3\sigma$ confidence interval of GPR.}
	\label{fig:GPfit}
\end{figure*}

\begin{table*}
\renewcommand{\arraystretch}{2}
\caption{GPR results: the peak time ($T_\mathrm{m}$, MJD) and flux density ($F_\mathrm{m}$, Jy) of the flares at 4.8\,GHz to 37\,GHz. The values are median, 84/16 percentiles (and median absolute deviation in brackets).}
\label{tab:gp}

{\centering	
	\begin{tabular}{|c|c|c|c|c|c|c}
	\hline
	&  & 37\,GHz & 22\,GHz & 15\,GHz & 8\,GHz & 4.8\,GHz \\ 
	\hline 
	\multirow{2}{*}{A1}  & $T_\mathrm{m}$ & \dots & \dots & $45030^{+2}_{-1} (1.4)$ & $45034^{+2}_{-2} (1.8)$ & $45057^{+6}_{-5} (5.9)$  \\ 
	& $F_\mathrm{m}$ & \dots & \dots & $4.53^{+0.04}_{-0.04}$ & $4.30^{+0.02}_{-0.03}$ & $3.65^{+0.07}_{-0.06}$ \\ 
	\multirow{2}{*}{B1}  & $T_\mathrm{m}$ & $46866^{+14}_{-9} (10.3)$ & $46857^{+5}_{-17} (8.8)$ & $46888^{+8}_{-3} (5.5)$ & $46900^{+3}_{-3} (2.9)$ & $46913^{+15}_{-8} (11.9)$ \\ 
	& $F_\mathrm{m}$ & $4.49^{+0.22}_{-0.14}$ & $4.15^{+0.21}_{-0.13}$ & $4.09^{+0.03}_{-0.04}$ & $4.01^{+0.05}_{-0.05}$ & $3.32^{+0.19}_{-0.21}$ \\ 
	\multirow{2}{*}{D1}  & $T_\mathrm{m}$ & $48904^{+4}_{-3} (3.4)$ & $48912^{+5}_{-4} (5.1)$ & $48903^{+1}_{-1} (0.9)$ & $48911^{+1}_{-1} (1.1)$ & $48930^{+1}_{-4} (2.2)$ \\ 
	& $F_\mathrm{m}$ & $6.70^{+0.13}_{-0.11}$ & $5.84^{+0.10}_{-0.09}$ & $6.11^{+0.04}_{-0.05}$ & $5.31^{+0.04}_{-0.04}$ & $4.17^{+0.02}_{-0.02}$ \\  
	\multirow{2}{*}{E1} & $T_\mathrm{m}$ & $50873^{+3}_{-3} (2.8)$ & $50876^{+3}_{-2} (2.4)$ & $50907^{+10}_{-8} (9.1)$ & \dots &  \dots \\ 
	& $F_\mathrm{m}$ & $5.30^{+0.13}_{-0.10}$ & $5.08^{+0.07}_{-0.07}$ & $4.01^{+0.07}_{-0.07}$ & \dots & \dots \\ 
	\multirow{2}{*}{E2}  & $T_\mathrm{m}$ & \dots & $51001^{+5}_{-8} (6.6)$ & $51015^{+4}_{-3} (3.8)$ & $51008^{+2}_{-2} (2.1)$ & $51027^{+7}_{-10} (8.0)$ \\
	& $F_\mathrm{m}$ & \dots & $4.80^{+0.14}_{-0.14}$ & $4.56^{+0.15}_{-0.15}$ & $4.18^{+0.04}_{-0.05}$ & $3.25^{+0.03}_{-0.05}$ \\ 
	\multirow{2}{*}{E3}  & $T_\mathrm{m}$ & \dots & \dots & $51492^{+1}_{-2} (1.6)$ & $51496^{+7}_{-5} (6.2)$ & $51530^{+3}_{-4} (3.7)$ \\ 
	& $F_\mathrm{m}$ & \dots & \dots & $2.30^{+0.02}_{-0.02}$ & $2.02^{+0.05}_{-0.04}$ & $1.66^{+0.02}_{-0.01}$ \\ 
	\multirow{2}{*}{F1}  & $T_\mathrm{m}$ & $53723^{+9}_{-13} (11.7)$ & \dots & $53752^{+7}_{-6} (6.3)$ & $53763^{+5}_{-4} (4.6)$ & $53752^{+1}_{-2} (1.5)$ \\ 
	& $F_\mathrm{m}$ & $3.42^{+0.12}_{-0.13}$ & \dots & $3.26^{+0.04}_{-0.04}$ & $2.62^{+0.04}_{-0.04}$ & $2.06^{+0.02}_{-0.01}$ \\ 
	\multirow{2}{*}{F2}  & $T_\mathrm{m}$ & $54203^{+4}_{-4} (3.7)$ & \dots & $54251^{+14}_{-12} (13.1)$ & $54269^{+6}_{-8} (7.3)$ & $54290^{+7}_{-6} (6.8)$ \\ 
	& $F_\mathrm{m}$ & $4.96^{+0.11}_{-0.11}$ & \dots & $2.78^{+0.14}_{-0.09}$ & $3.09^{+0.08}_{-0.07}$ & $2.39^{+0.09}_{-0.09}$ \\ 
	\multirow{2}{*}{G1}  & $T_\mathrm{m}$ & $54764^{+3}_{-2} (2.7)$ & \dots & $54770^{+1}_{-0} (1)$ & $54774^{+1}_{-2} (1.2)$ & $54784^{+2}_{-1} (1.9)$ \\ 
	& $F_\mathrm{m}$ & $6.00^{+0.12}_{-0.13}$ & \dots & $6.48^{+0.03}_{-0.03}$ & $6.57^{+0.05}_{-0.05}$ & $5.37^{+0.06}_{-0.07}$ \\ 
		\hline 

	\end{tabular} 
}

\end{table*}

The hyperparameters are inferred by maximizing the following marginal likelihood function:
\begin{equation}
    \mathcal{L} = \log p(\mathbf{y|t,\Theta}) = -\frac{1}{2}\mathbf{y}^TC^{-1}\mathbf{y} - \frac{1}{2}\log |C| - \frac{n}{2}\log2\pi,
\end{equation}
where $\mathbf{\Theta}$ is vector of the hyperparameters, and $p(\mathbf{y|t,\Theta})$ is the probability of realization of the observed flux density values $\mathbf{y}={\{y_1, \dots, y_n\}}$ at the times $\mathbf{t}={\{t_1,\dots, t_n\}}$ for the given $\mathbf{\Theta}={\{A, l, \epsilon, \sigma_n\}}$. 
Then, using the obtained hyperparameters and the data at hand one can obtain the mean value of the GP realizations and its corresponding confidence interval at any time point $t_k$, i. e. obtain the regression model for the light curve with the uncertainties~\citep{Rasmussen:GPM}.

The GPR constructed as described above, however, does not account for the uncertainties of the estimated hyperparameters. This can be incorporated by sampling the posterior distributions of the hyperparameters with Markov chain Monte Carlo (MCMC) method. For that purpose we employed \texttt{emcee} sampler~\citep{2013PASP..125..306F} and \texttt{george} python library \citep{2015ITPAM..38..252A}. We used uniform priors for the hyperparameters. 
From the obtained samples of the joint posterior distributions we drew a number of the hyperparameters vectors $\mathbf{\Theta}$ and constructed the realizations of the GP. For each realization we measured time and flux density of the flares peaks and used their scatter for the uncertainties estimation.

To check whether there is a difference between time lags at different epochs we split our data manually into seven time intervals each having at least one prominent outburst. This was done to perform cross-correlation analysis (see below). The selected intervals are denoted in Figure~\ref{fig:lightcurves}. 
The results are summarized in Table~\ref{tab:gp}. 
As an example, the results of implementing GPR to the 15 GHz light curve of epochs interval 1985 -- 1988 (labeled B) are shown in Figure~\ref{fig:GPfit}. 
The histograms illustrate the resulting posterior distributions of the hyperparameters (the plot is produced with \texttt{corner} package by~\citealt{corner}). Median and 16/84 percentiles are shown by vertical lines. In the upper right corner the light curve is shown along with $\pm3\sigma$ (99.7\,\%) confidence interval for GPR. Note the outlying data point near the flare peak. Its sharpness and absence of the counterparts at other frequencies suggest its ``extrinsic'' nature presumably due to measurements error. The outlier is effectively bypassed by the GPR method, which indicates the low probability of it to belong to the model. The hyperparameter $\sigma_n$ shows the level of an additional unaccounted noise in data at the level about 0.1 -- 0.2\,Jy, which is comparable with the data uncertainties. Thus, using the jitter in our kernel provides more realistic uncertainties of the $T_\mathrm{m}$ and $F_\mathrm{m}$.
As expected, the confidence interval widens within data gaps. The separate flares and the corresponding time delays are shown in Figure~\ref{fig:flares-lags}.  

\begin{figure}
\includegraphics[width=\columnwidth,trim=0cm 1cm 0cm 0cm]{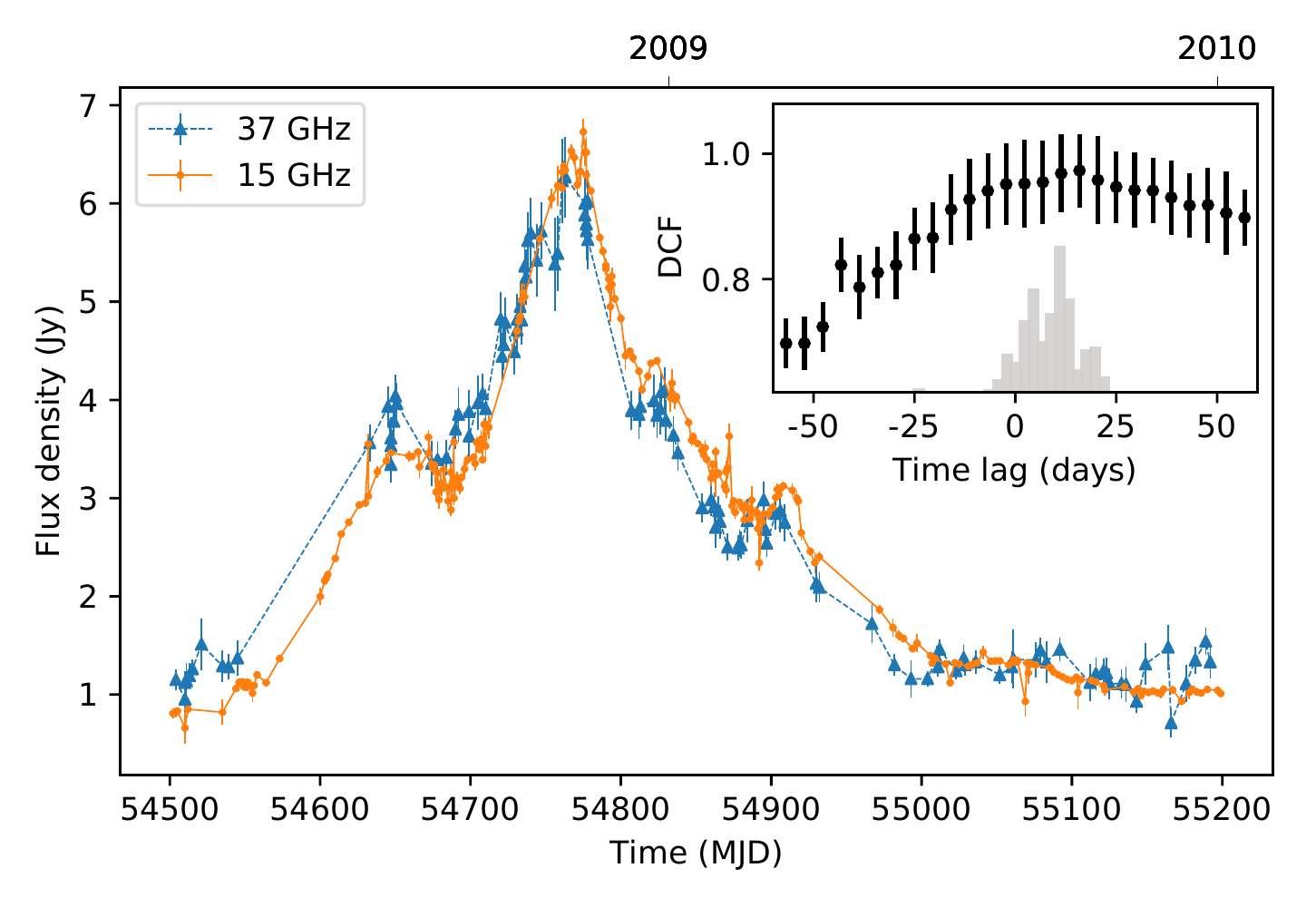}
\caption{The G-flare at 37 and 15 GHz. Their mutual DCF is shown in the upper right exes. The time lags distribution  obtained with FR/RSS simulations are shown with grey bars (not normalized). The resulting marginal time delay is $\Delta T = 12\pm 8$\,days.}
\label{fig:DCF}
\end{figure}

We also apply the DCF method to the time intervals labeled in Figure~\ref{fig:lightcurves}. To estimate the errors of the resulting delays we perform Monte Carlo simulations and modified bootstrapping (or flux randomization / random sub-sample selection (FR/RSS) method proposed by \citealt{1998PASP..110..660P}). This allows us to account for errors due to initial flux-measurement uncertainties as well as for the data outliers. To account for the possible influence of the time bin selection we varied it in each simulation to be uniformly spread in the range $[0.5, 1.5]\Delta t_\mathrm{mean}$ (where $\Delta t_\mathrm{mean}$ is the mean time span between the observations. Example of the DCF results for the G-flare at 37--15\,GHz is shown in Figure~\ref{fig:DCF} along with the obtained distribution of the time lags. In general, the DCF results are similar to that by GPR but the wide resulting distributions of the time delays give higher uncertainties. Moreover, the DCF results are affected by the selection of a light curve length. The following discussion is based on the results obtained with the GPR method.


\section{Variability time delays}
\label{sec:lags}

\begin{figure*}
    \includegraphics[width=\linewidth]{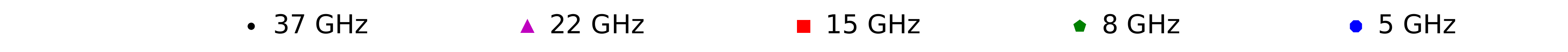}\\
	\includegraphics[width=0.55\linewidth,trim=1.2cm 0.5cm 1.2cm 0cm]{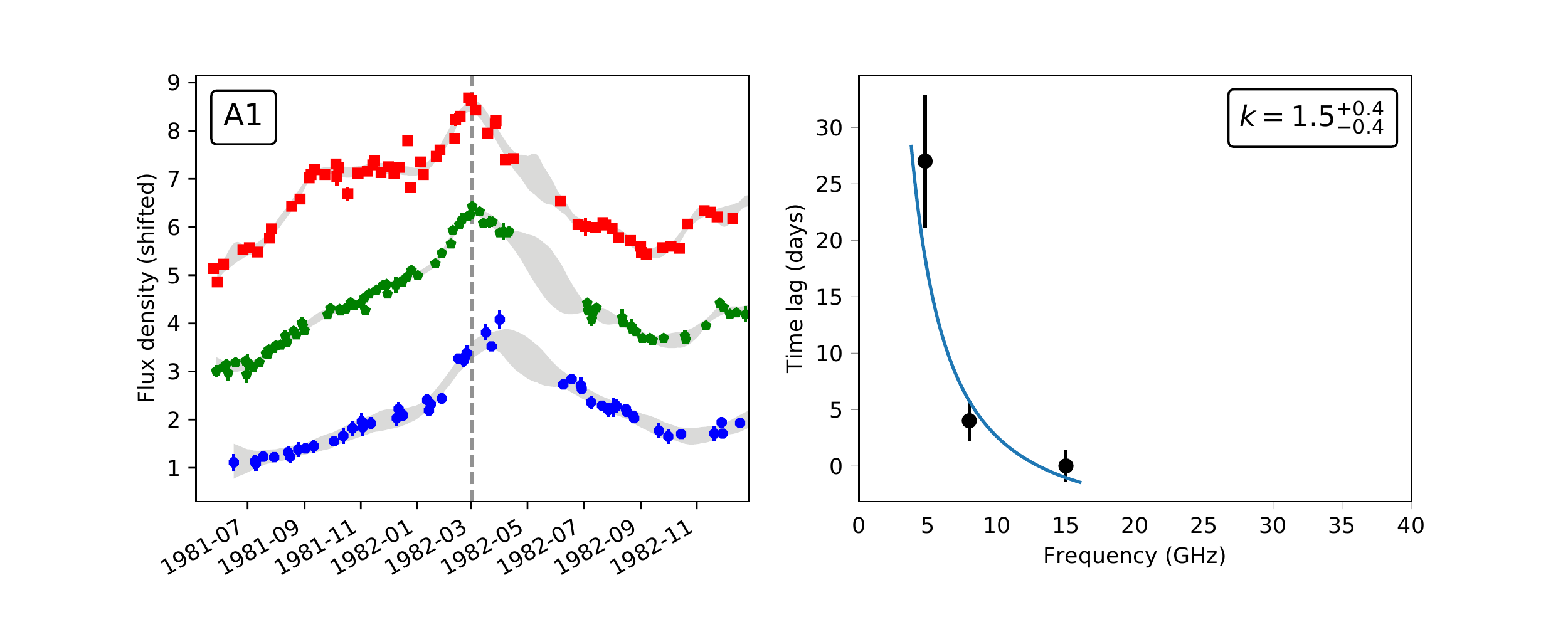}\includegraphics[width=0.55\linewidth,trim=1.2cm 0.5cm 0cm 0cm]{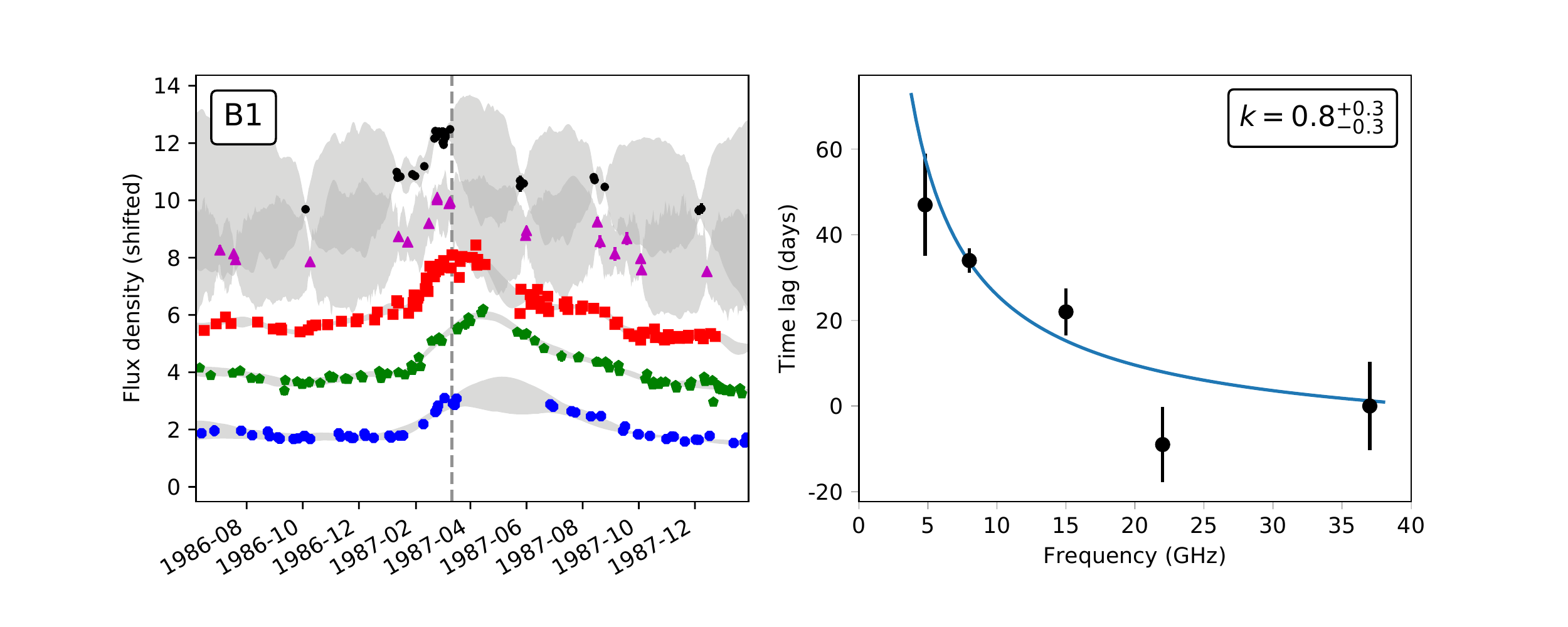}
	\includegraphics[width=0.55\textwidth,trim=1.2cm 0.5cm 1.2cm 0cm]{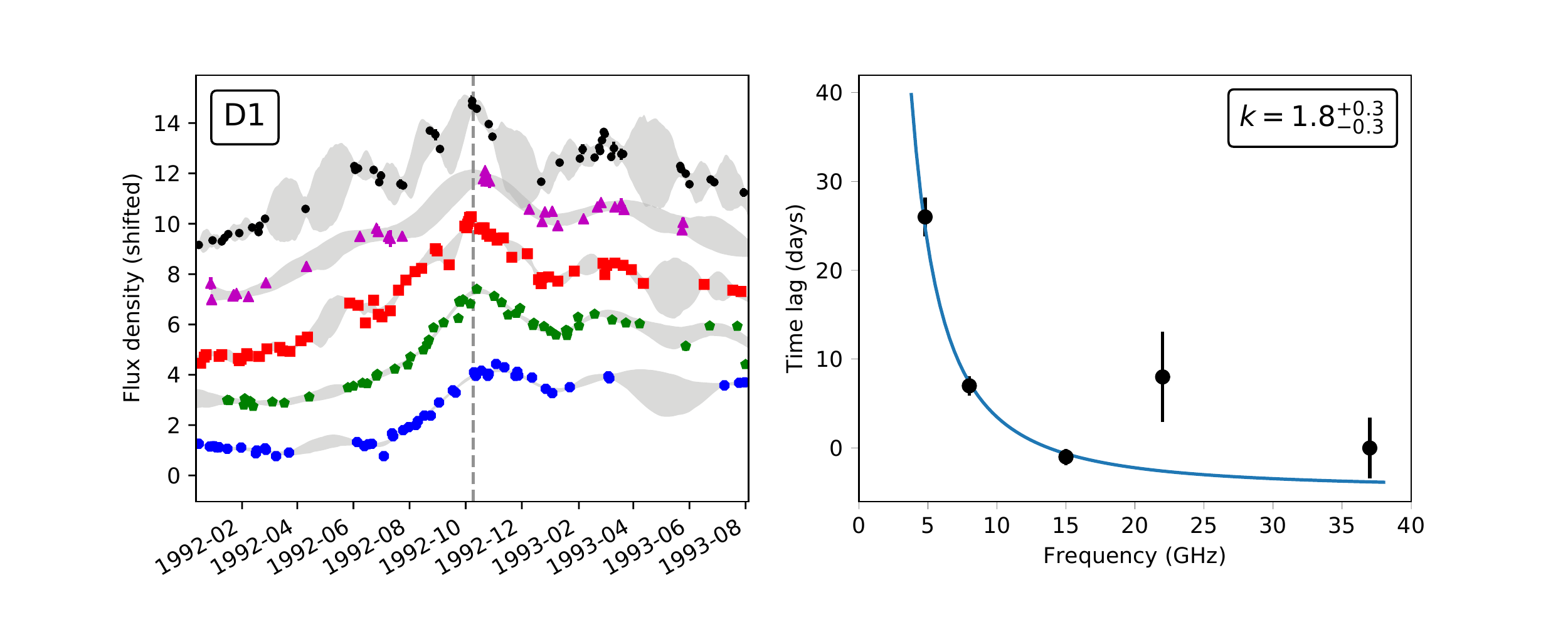}\includegraphics[width=0.55\textwidth,trim=1.2cm 0.5cm 0cm 0cm]{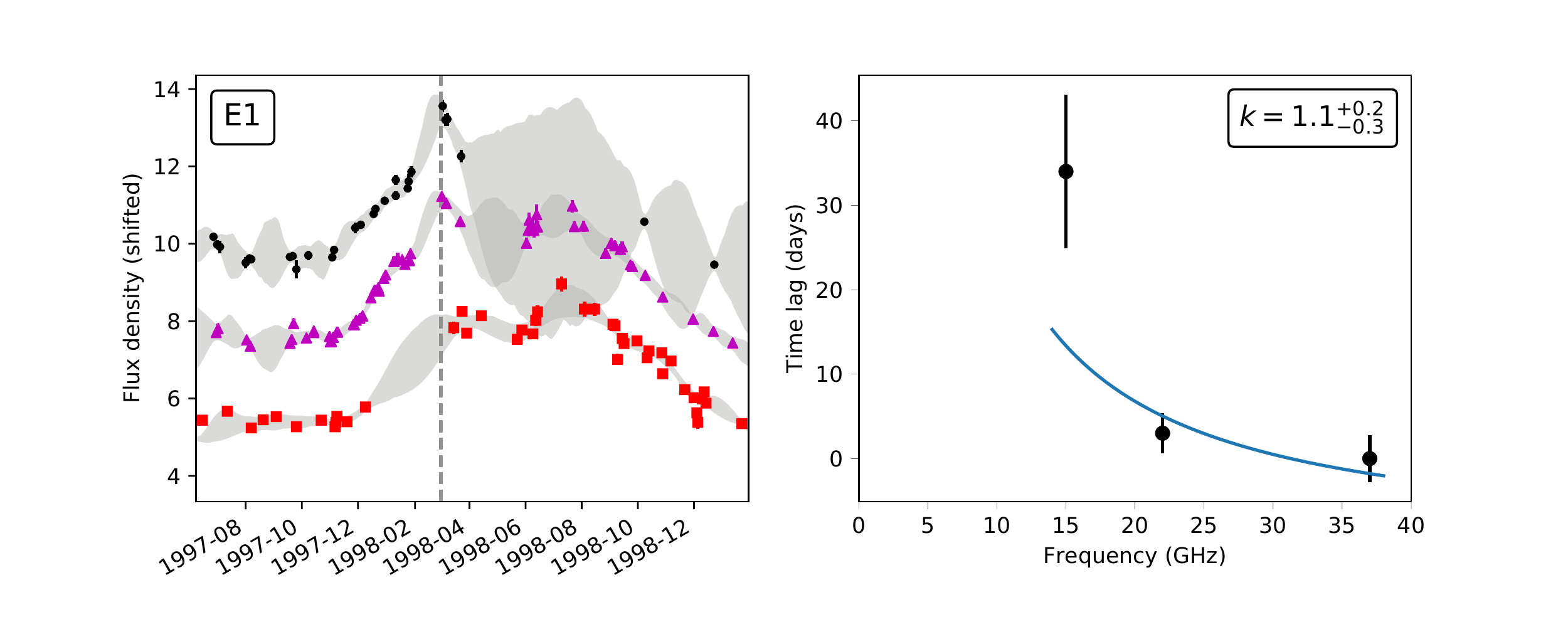}
	\includegraphics[width=0.55\textwidth,trim=1.2cm 0.5cm 1.2cm 0cm]{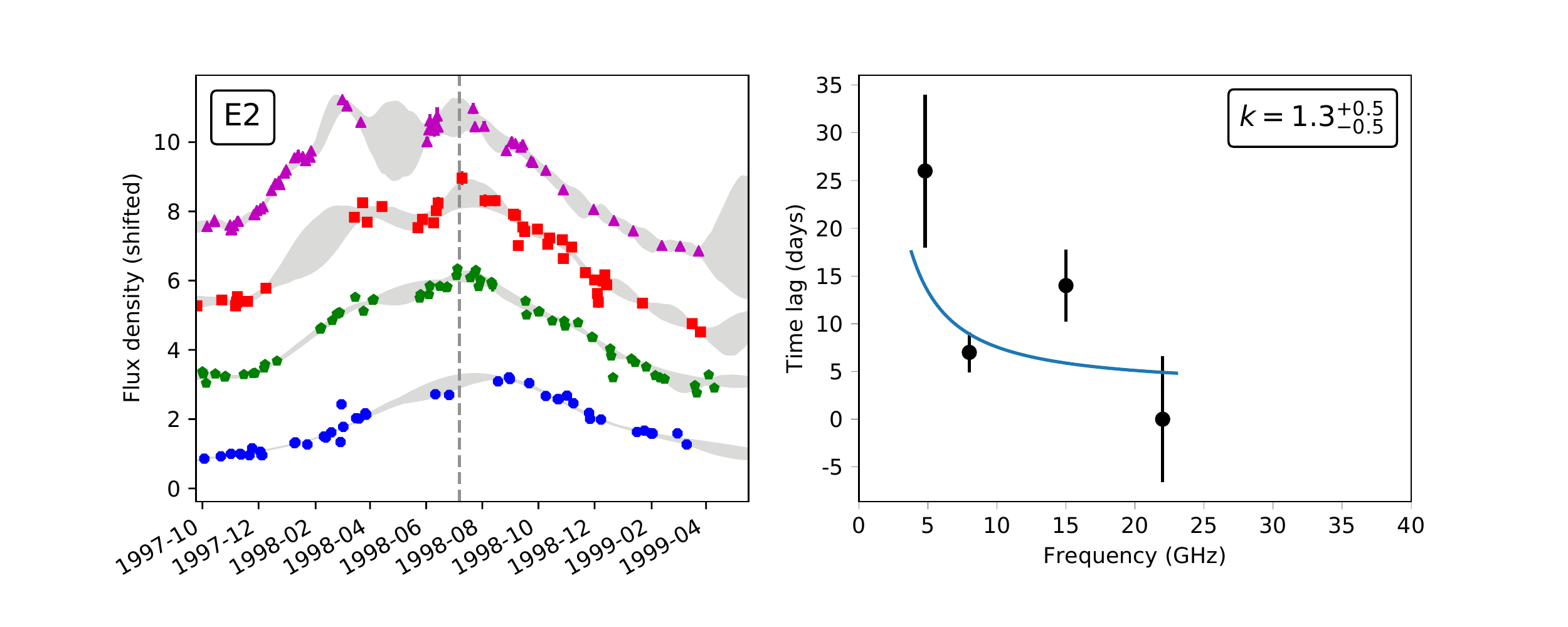}\includegraphics[width=0.55\textwidth,trim=1.2cm 0.5cm 0cm 0cm]{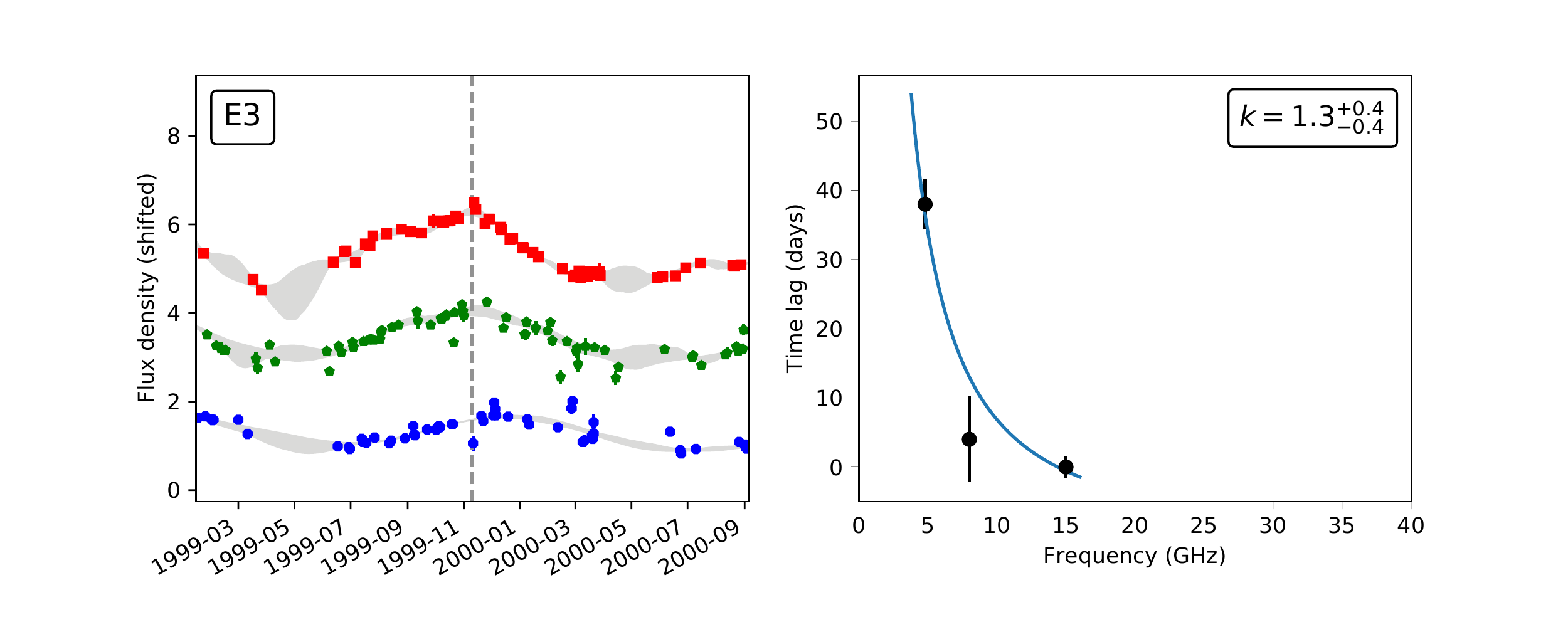}
	\includegraphics[width=0.55\textwidth,trim=1.2cm 0.5cm 1.2cm 0cm]{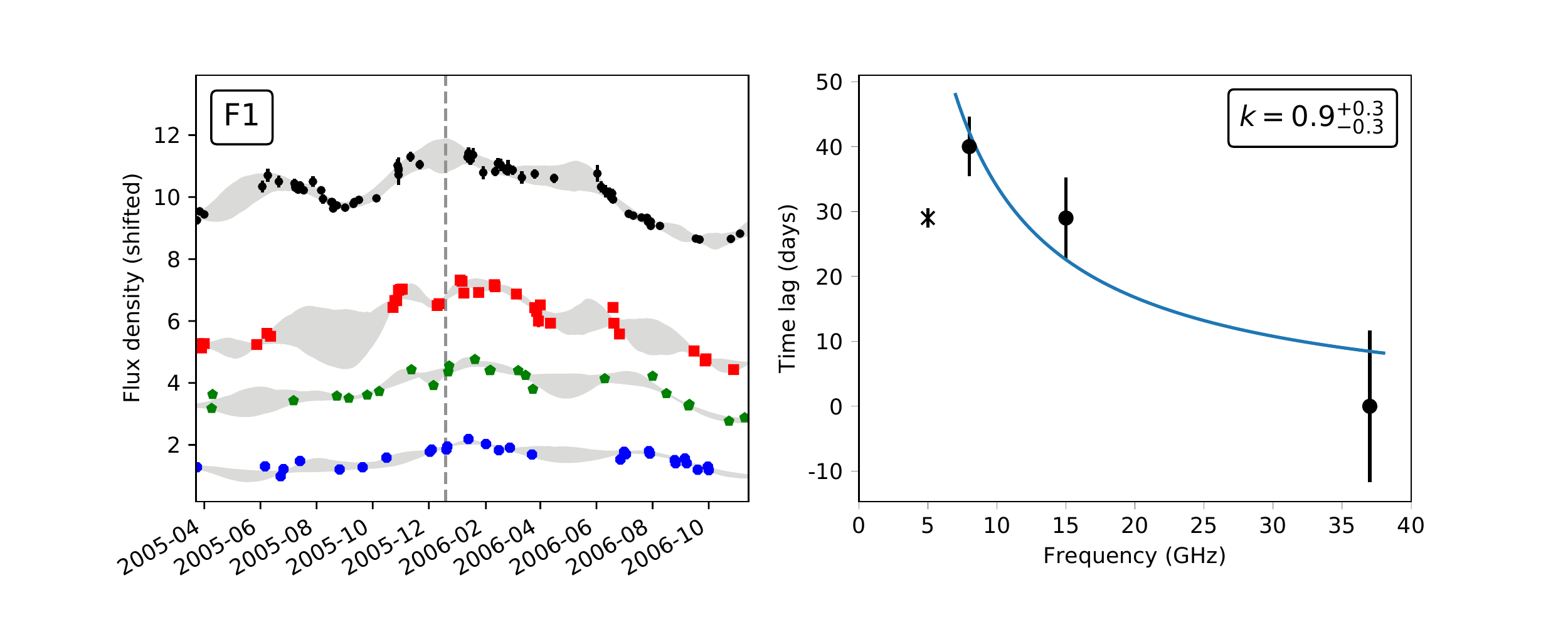}\includegraphics[width=0.55\textwidth,trim=1.2cm 0.5cm 0cm 0cm]{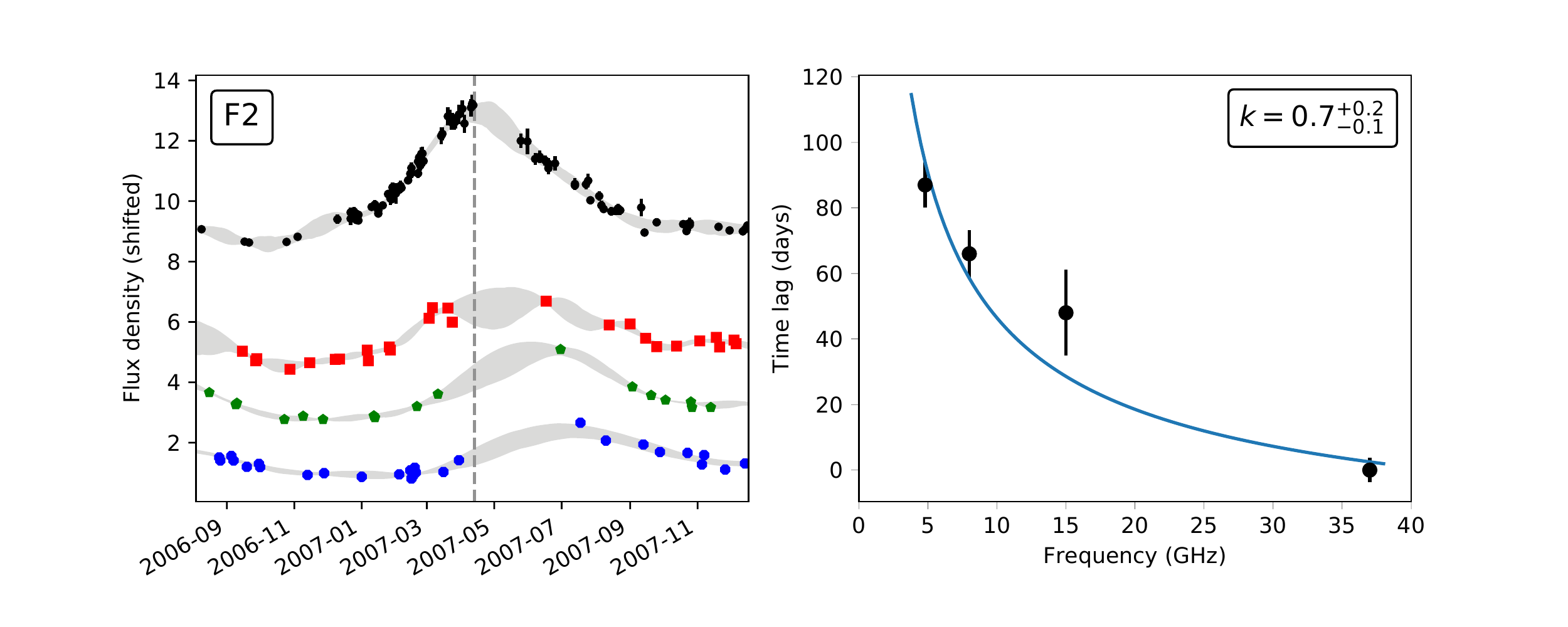}
	\includegraphics[width=0.55\textwidth,trim=1.2cm 1cm 1cm 0cm]{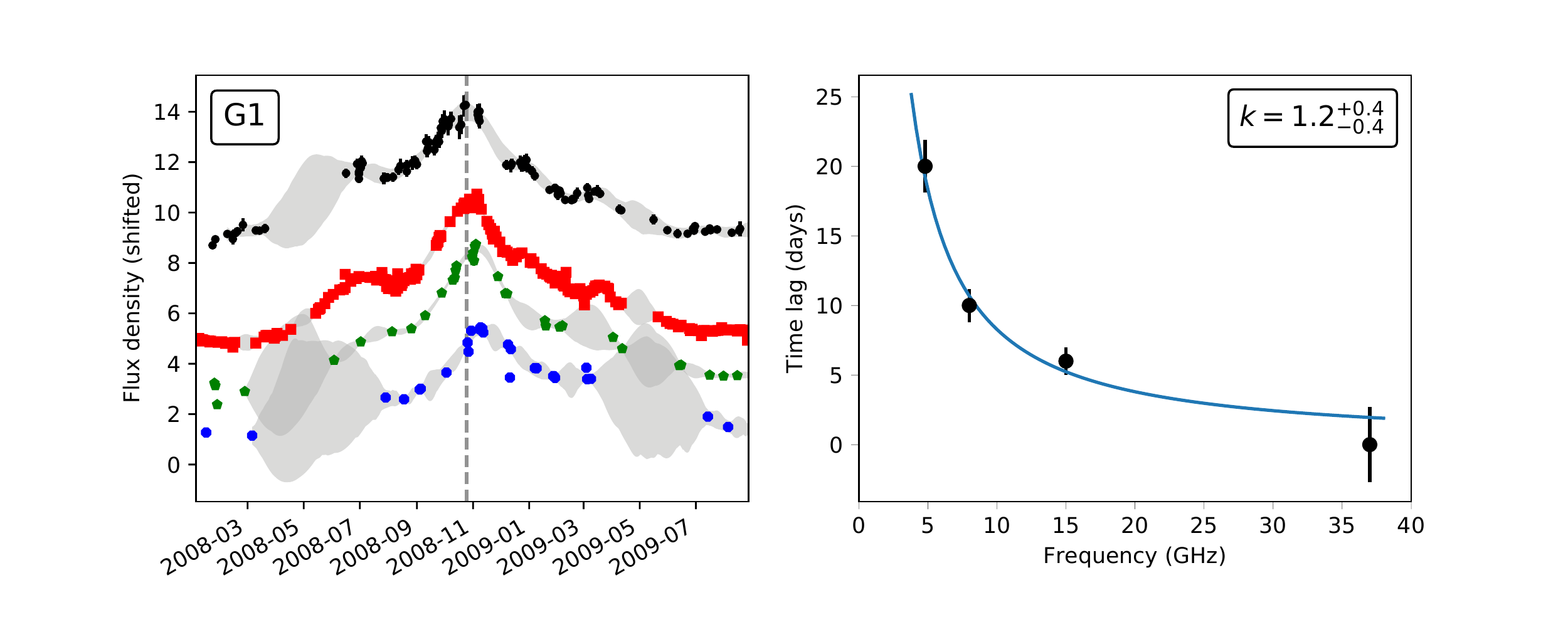}\includegraphics[width=0.52\textwidth,trim=1cm 1cm 0cm 0cm]{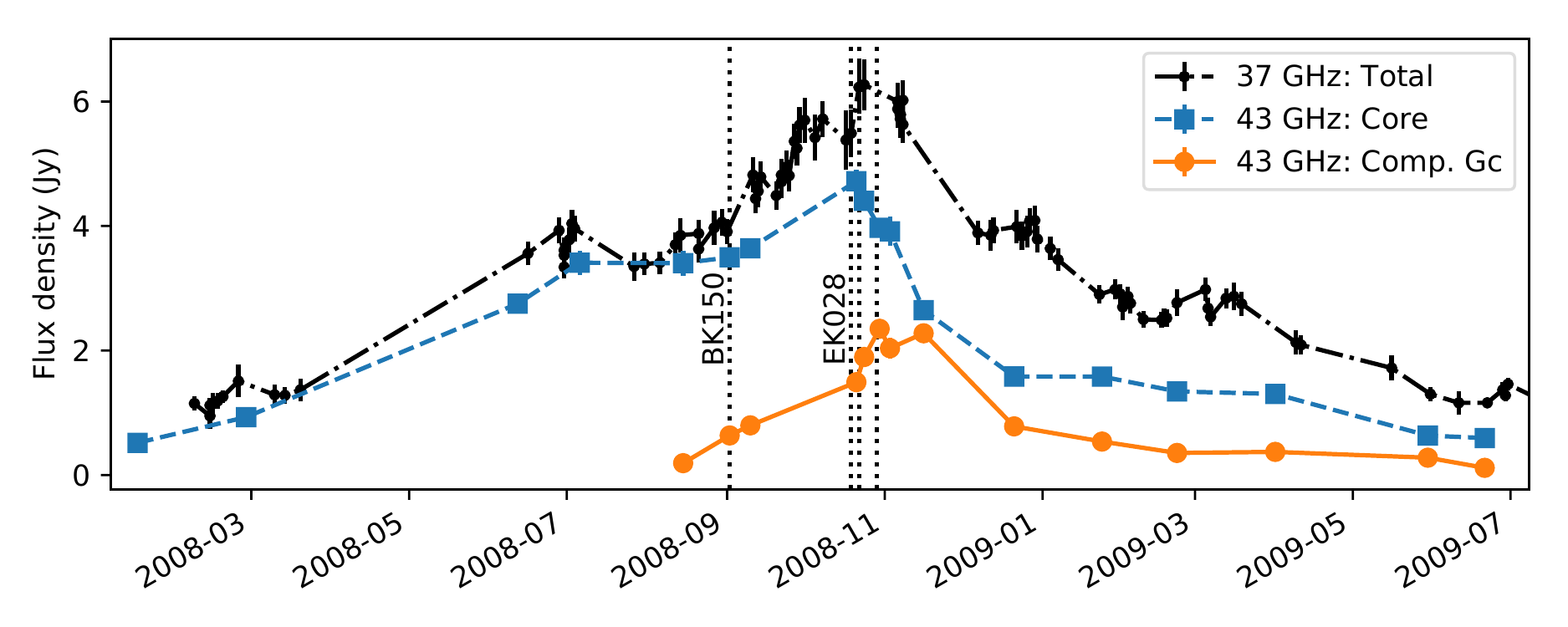}
	\caption{Separate flares and corresponding time lags. Vertical dashed lines mark the peak time at the highest frequency. The grey shaded area corresponds to $\pm 3\sigma$ confidence interval of the GPR. The light curves of single-dish 37\,GHz, the core, and the component Gc for the G-flare are shown on the last panel (the multi-frequency VLBI experiments are marked with vertical dotted lines). 
	}
	\label{fig:flares-lags}
	
\end{figure*}

We approximated the dependence of the time lags $\Delta T$ on frequency $\nu$ (comparative to the peak at highest one) with the model $\Delta T = a_t(\nu/GHz)^{-k_t} + b_t$. 
To find the parameters we used the \texttt{emcee} MCMC sampler with uniform prior distributions of $a_t$ and $b_t$ and normal prior distribution of $k_t^{prior} = 0.9\pm0.44$. This estimate was obtained by \citealt{2015MNRAS.452.4274P} for the ''core size -- frequency`` dependence for a large AGN sample and is a reasonable guess in case of a conical non-accelerated jet (e.g., \citealt{2014MNRAS.437.3396K} showed that the core size and the time lags follow the same power-law in the blazar 3C\,454.3). 

The parameter $b_t < 0$\,days corresponds to the delay between the flare peak at a given highest frequency and ''infinite`` frequency, $a_t>0$\,days characterizes the flow speed, and $k_t$ reflects the opacity mechanism.
As seen in Figure~\ref{fig:flares-lags} the measured peak-to-peak delays differ from one flare to another. Moreover, there are flares with unusual lags when a peak comes earlier at lower frequency. (Table~\ref{tab:gp}). 
Similar unexpected time delays in the source have been reported by \cite{1999A&A...344..807K} for a short flare in 1992 that is a part of D-flare in our notation.

\begin{figure}
\includegraphics[width=\columnwidth,trim=0cm 1cm 0cm 0cm]{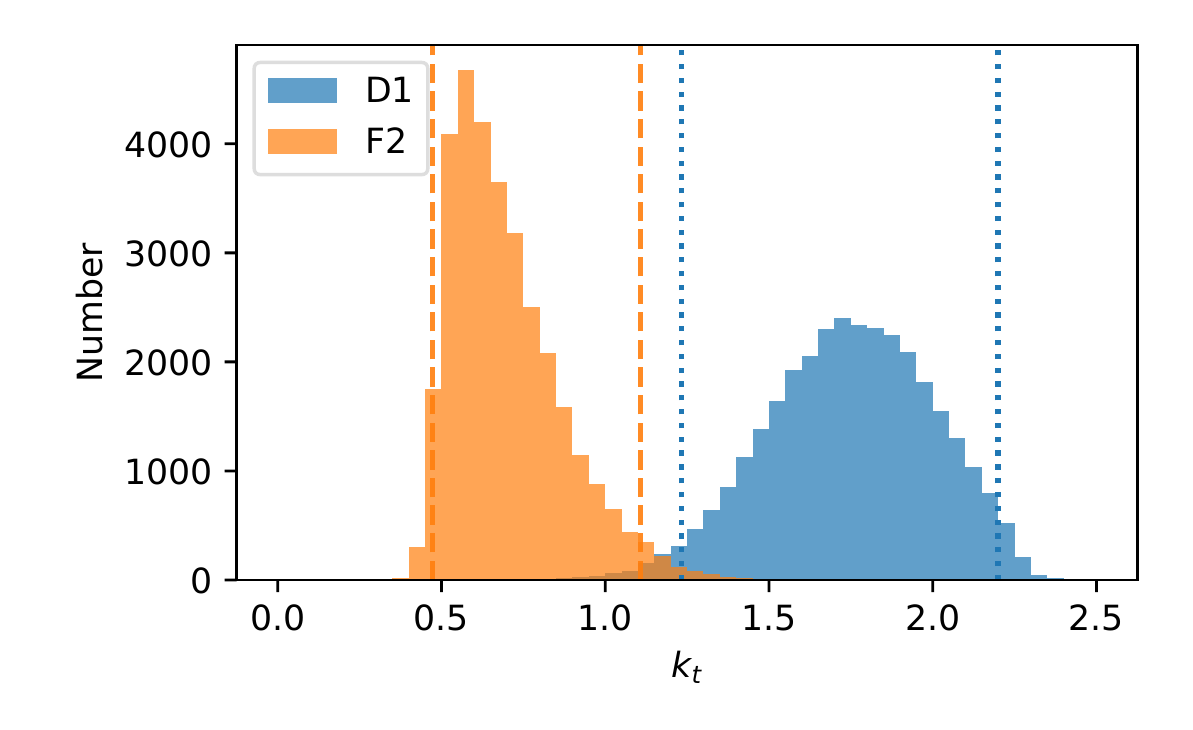}	
\caption{Posterior distributions of $ k_t $ parameter for D1 and F2 flares. Vertical dotted and dashed lines denote the 95\% intervals, indicating significant difference of $k_t$.}	
\label{fig:k_distributions}
\end{figure}

We obtain a wide range of parameters $0.7 \le k_t \le 1.8$ with median value $k_t=1.2$. The posterior $k_t$ distributions are plotted in Figure~\ref{fig:k_distributions} for the two marginal cases (D1-flare with $k_t=1.8$ and F2-flare with $k_t=0.7$), showing the difference between the $k_t$ values with confidence more than 95\%.
We find no correlation between $k_t$ and the peak flux density of the flares. The changing of $k_t$ in time is not monotonic. 

Different values of the index $k_t$ within one source for separate flares have been also reported by \cite{2011MNRAS.415.1631K}. The reason for its change is not obvious, and may be related to the intrinsic changes in the jet, e.g when the injected particles distort the profile of initial electron distribution in the outflow. 

In the last panel of Figure~\ref{fig:flares-lags} the light curve of the 37\,GHz single-dish flux density is shown along with the 43\,GHz light curves of the core and the superluminal component \textit{Gc} (see also Section~\ref{sec:component}). Peak-to-peak time delay between the component and the core flares obtained using GPR is $T_{Gc, max} - T_{Core, max} = 21\pm4$ days. One can see, that the total flux density measured by the single-dish telescopes is the sum of the core and Gc and has a maximum somewhere in the middle between their peaks. Therefore, the measured time lags between the single-dish flares might be biased (i.e. systematically differ from the delay between the moments of passing of a disturbance through the core). The flare profile of the total flux density differs from that of the core since the former is the superposition of the latter and the one of the component Gc. {While the rising parts of the profiles are similar, the peak and decay differ: they are smoother for the total flux density profile, as can be seen in the last panel of Figure~\ref{fig:flares-lags}}. Therefore, if a single-dish light curve is used for estimation of brightness temperature and Doppler factor, there might be a bias due to overestimation of the variability time-scale (see also section~\ref{sec:tb}).


\section{Core size and spectrum}
\label{sec:core}

The frequency dependence of the apparent core size follows power law $W_{core}\propto \nu^{-k_w}$, where $k_w$ characterizes the opacity mechanism. The index $k_w\approx1$ when the synchrotron self-absorption dominates, and $k_w \gtsim{2}$ in case of prevailing free-free absorption~\citep[e.g., ][]{2008arXiv0811.2926Y}.  
\citet{2015MNRAS.452.4274P} found $k_w = 0.90\pm0.44$ based on quasi-simultaneous multi-frequency observations of a large AGN sample. This value was measured for the sources located at the Galactic latitudes higher than $\sim10^\circ$. The authors also found $k_w\approx1.8$ for the Galactic plane residents indicating significant interstellar scattering.

\begin{figure}
	\centering
	\includegraphics[width=\columnwidth,trim=0cm 1cm 0cm 0cm]{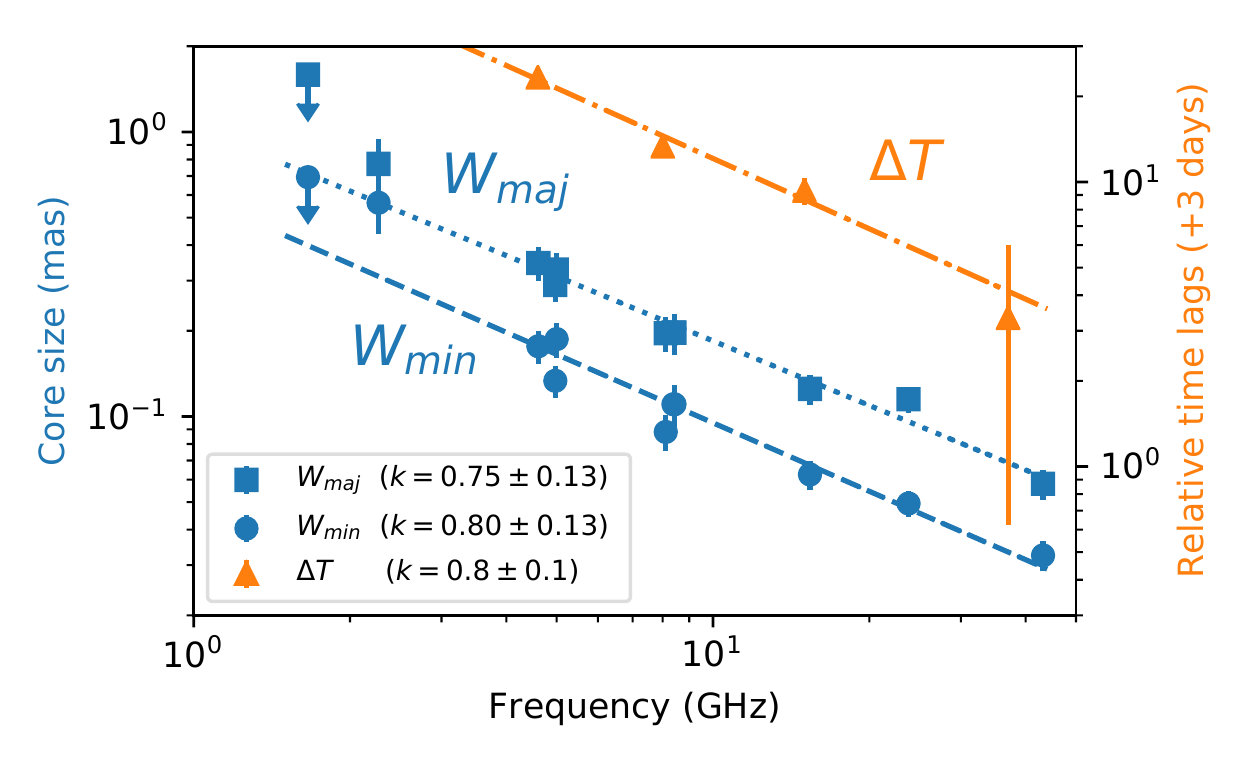}
	\caption{The core size $W$ (left axis) and the time lags of the G-flare peaks $\Delta T$ (right axis) vs.\ observations frequency.}
	\label{fig:coresize_fit}
\end{figure}

The core in both multi-frequency VLBI experiments is well described by the elliptical Gaussian model with the mean axial ratio $\epsilon \sim 0.5$ (see Table~\ref{tab:difmap_models}). The dependence of its major and minor axes (full width at half-maximum (FWHM) of the Gaussian) on frequency is approximated with the model $W(\nu) = a_w \nu^{-k_w}$ (an absence of an additional constant term here suggests that the core has ``zero'' size at ``infinite'' frequency). Since there is no significant difference between the core size obtained in both VLBI experiments at 5\,GHz and 8\,GHz, we merge all the data from these experiments for a better fit representation (see Figure~\ref{fig:coresize_fit}).

Both major and minor axes follow the power law with $k_{w, maj} = k_{w, min} = 0.8\pm0.1$ ($a_{w, maj} = 1.1\pm0.1$, $a_{w,min} = 0.65\pm0.07$), which is in good agreement with the photosphere scenario.
Since the modeled core size at 1.7\,GHz does not fulfill the resolution criteria, the upper limit is used. The core at 2.3\,GHz is formally resolved, however, its size $W_{maj}\approx0.8$\,mas is larger than that expected within the synchrotron self absorption model ($\sim0.6$\,mas). The axial ratio $e>0.7$ at this band also falls out from the rest data. The interstellar scattering might be significant at these frequencies. For the given celestial position the angular broadening predicted by the \texttt{NE\,2001} model~\citep{2002astro.ph..7156C} is $W_\mathrm{NE, 2.3\,GHz} = 0.16$\,mas, much lower than the core size obtained from modeling. This suggests that if the scattering comes into play here, there is really much denser Galactic medium in that direction, than proposed by the \texttt{NE\,2001} model. We also note, that excluding the data at 1.7\,GHz and 2.3\,GHz does not affect the results of the fit.

The power-law index found for the core size dependence $k_w$ is measured with accuracy better than $k_t$ and can be used to constrain the model-fit parameters of the time lags dependence for the G-flare (within the assumption of a conical non-accelerated jet). For that purpose we employ the \texttt{emcee} sampler with narrow (informative) prior distribution for $k_t = k_w = 0.8\pm0.1$ and obtain distributions of the other two parameters, yielding $a_t = 80\pm14$ days, and $b_t = -3\pm2$ days. The time lags and the refined model are shown in Figure~\ref{fig:coresize_fit} (the right axis).

\begin{figure}
	\includegraphics[width=\columnwidth,trim=0cm 1cm 0cm 0cm]{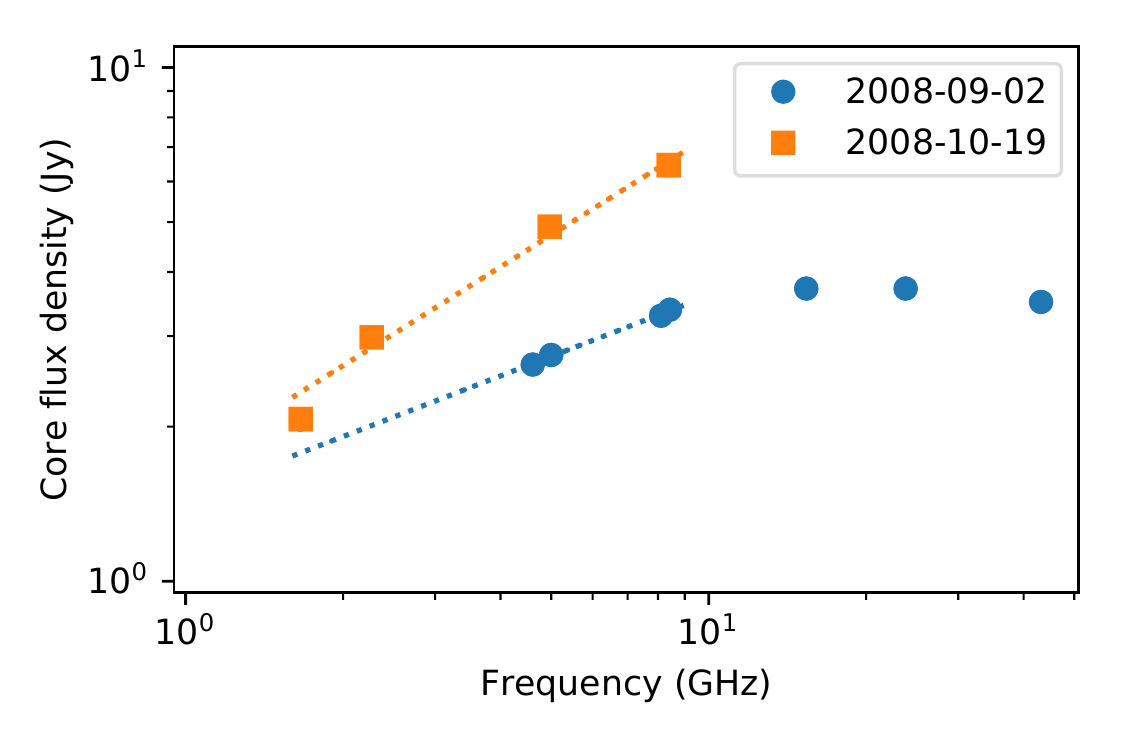}
	\caption{The core spectra for two epochs during the flare measured from VLBA and EVN data.}
	\label{fig:corespec}
\end{figure}

The core spectrum is shown in Figure~\ref{fig:corespec} for two multi-frequency experiments (the core flux density at a given band is assumed to remain constant during each experiment and a possible shift of the core is neglected). The optically thick spectral index increases from $\alpha_1 = 0.39\pm0.09$ to $\alpha_2 = 0.64\pm0.05$, {possibly due to the peak at lower frequencies lagging with respect to higher frequencies}. We also discuss possible influence of adiabatic losses in Section~\ref{sec:coreshift}.

\section{Brightness temperature and Doppler factor}
\label{sec:tb}

\subsection{Ground-based data}

The apparent brightness temperature of a jet component in the source frame can be estimated using the model parameters (e.g., \citealt{2005AJ....130.2473K}):

\begin{equation}
\label{eq:Tb}
T_\mathrm{b} = 1.22\times10^{12}\mathrm{K}\,\,\frac{S/\mathrm{Jy}\,\, (1+z)}{(\nu/\mathrm{GHz})^2\, W_\mathrm{maj}/\mathrm{mas}\,\,W_\mathrm{min}/\mathrm{mas}},
\end{equation}
where $S$ -- flux density, $W_\mathrm{maj}$ and $W_\mathrm{min}$ -- major and minor apparent size (FWHM of a Gaussian model), $z$ -- redshift, and $\nu$ -- observing frequency. 

On the other hand it is possible to estimate the so-called variability brightness temperature in the source frame assuming that the variability time-scale $t_\mathrm{var}$ corresponds to the light-crossing time of the component size (e.g., \citealt{2009A&A...494..527H}):

\begin{equation}
T_\mathrm{b,var} = 1.05\times10^8 \mathrm{K}\,\, \frac{(D_\ell/\mathrm{Mpc})^2\, S_\mathrm{obs}/\mathrm{Jy}}{(1+z)(\nu/\mathrm{GHz})^2(t_\mathrm{var}/\mathrm{yr})^2},
\label{eq:tbvar}
\end{equation}
where $D_\ell$ -- luminosity distance, $S_\mathrm{obs}$ -- flux density, $\nu$ -- observations frequency, $t_\mathrm{var}$ -- variability time-scale. 

$T_\mathrm{b}$ and $T_\mathrm{var}$ are measured experimentally and can be expressed through an intrinsic brightness temperature $T_\mathrm{int}$, amplified by Doppler boosting (e.g., \citealt{2007Ap&SS.311..231K}):

\begin{equation}
\label{eq:Tb_sys}
\begin{cases}
T_\mathrm{b} = \delta T_\mathrm{int} \\
T_\mathrm{b,var} = \delta^3 T_\mathrm{int},
\end{cases}
\end{equation}
where $\delta$ is Doppler factor. From eq.~(\ref{eq:Tb_sys}) we obtain $\delta$ and $T_\mathrm{int}$.

We adopt the following parameters: $D_\ell = 6142$\,Mpc, $z=0.94$, $S_\mathrm{obs} = 4.7$\,Jy -- peak value of the core flux density during the G-flare, and $t_\mathrm{var} = 0.09$\,yr -- timescale of the core variability during the flare. Note, that the flare profile has a plateau and differs from the two-sided exponential. Moreover, the flare decays faster than it rises. We estimate the scale using the time interval and flux density change between the observations $i, j$ as $t_\mathrm{var} = |(t_{i} - t_{j})/\ln(S_{i}/S_{j})|$. The shortest time-scale is the pairs $(i,j)$ falls on the flare decay (epoch 2008-11-03). 
Thus, we use a self-consistent system to estimate Doppler factor and intrinsic brightness temperature, considering the size and variability timescale of the same region (the core).

We obtain $T_\mathrm{b,var} = 6.4\times10^{14}$\,K and $T_\mathrm{b} = 1.5\times10^{12}$\,K, which yields $\delta \approx 21$, and $T_\mathrm{int} \approx 7\times10^{10}$\,K. This value is close to the equipartition brightness temperature $T_\mathrm{eq} \simeq 5\times10^{10}$\,K, suggested by \cite{1994ApJ...426...51R}. 

\begin{figure*}
	\includegraphics[width=\textwidth,trim=0cm 1cm 0cm 0cm]{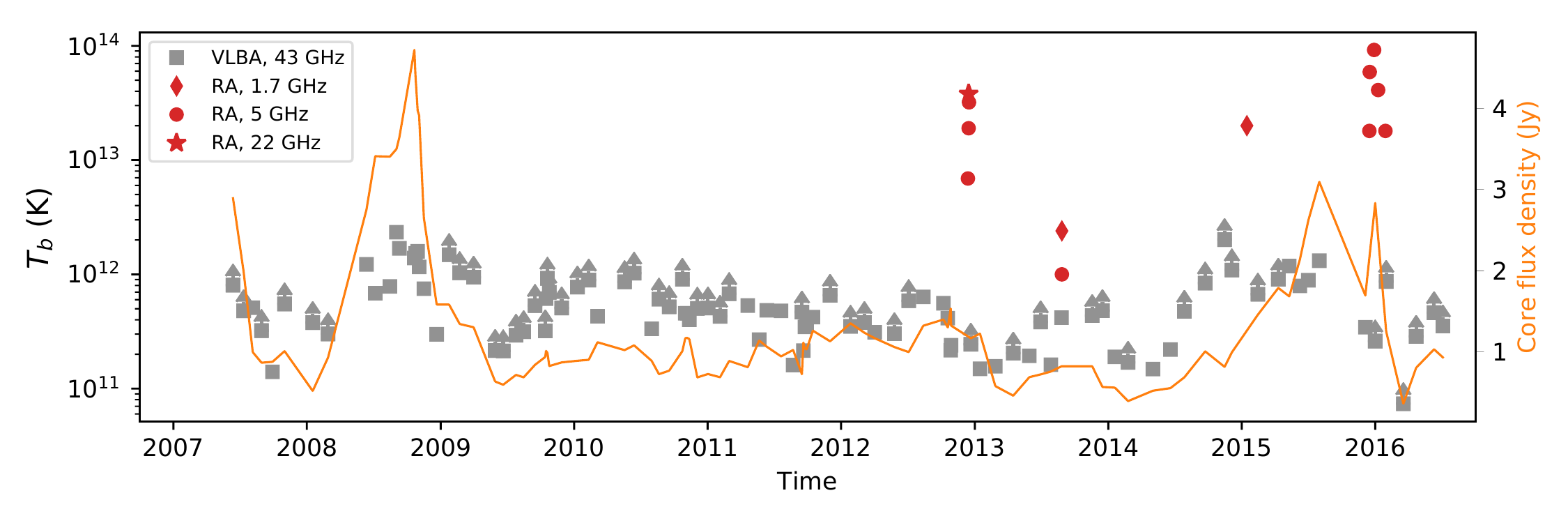}
	\caption{The apparent brightness temperature by VLBA at 43\,GHz and \textit{RadioAstron} at 1.7, 5 and 22\,GHz. Arrows mark $T_\mathrm{b}$ estimated using the resolution limits. The curve shows core flux density at 43\,GHz (right axis).} 
	\label{fig:Tb_core_flux}
\end{figure*}

We also estimate the apparent brightness temperature for each VLBA observation at 43\,GHz. These measurements are plotted with gray squares in Figure~\ref{fig:Tb_core_flux} (arrows indicate the lower limits of $T_\mathrm{b}$ in cases of unresolved core). Note that during the flares (2008 and 2015) the apparent brightness temperature increases by a factor of about 2.

\subsection{The RadioAstron data}
\label{sec:RA}

We estimate a lower limit of apparent brightness temperature $T_\mathrm{b}^\mathrm{min}$ in the source frame assuming that the brightness distribution has a circular Gaussian profile \citep{2015A&A...574A..84L}:

\begin{equation}
T_\mathrm{b}^\mathrm{min} = \frac{\pi e}{2k_B} L^2 V_L \approx 3.09 \mathrm{K} \left(\frac{L}{\mathrm{km}}\right)^2\left(\frac{V_L}{\mathrm{mJy}}\right) (1+z),
\end{equation}
where $k_\mathrm{B}$ is Boltzmann constant, $L$ -- projected baseline, $V_L$ -- visibility amplitude measured on that baseline. 
Uncertainty of the visibility amplitude is estimated from the statistical (thermal) noise of data as well as amplitude calibration errors. For the latter we adopt the typical 10\,\% value \citep{2014CosRe..52..393K}.
$T_\mathrm{b}^\mathrm{min}$ must be considered as the most conservative limit, which can be achieved for the given visibility amplitude.  
Further, we derive the size $W$ from modeling visibility amplitudes on ground-ground and ground-space baselines with a circular Gaussian profile and estimate the corresponding brightness temperature $T_\mathrm{b}^\mathrm{Gaus}$ (see eq.~(1,2) in~\citealt{2015A&A...574A..84L}). We note that $T_\mathrm{b}^\mathrm{Gaus}$ might  also be treated as a less strict lower limit, since the real component size can not exceed the estimate $W$ (see below). The lower limit $T_\mathrm{b}^\mathrm{min}$, the estimate $T_\mathrm{b}^\mathrm{Gaus}$ and the size $W$ are shown in columns 5-6 in Table~\ref{tab:ra}. 

\begin{table}
	\renewcommand{\arraystretch}{1.5}
	\caption{\textit{RadioAstron} observations of 0235+164. Columns are: (1) Date, (2) Band, (3) Projected baseline (in Earth diameters), (4) Lower limit of brightness temperature, (5) Estimate of the apparent brightness temperature in the source frame, (6) Estimate of circular Gaussian component size.} 	
	\label{tab:ra}		
	\begin{tabular}{|{c}|{c}|{c}|{c}|{c}|{c}|}
		\hline
		Date & $\nu$ & $D/D_\oplus$ & $T_\mathrm{b}^\mathrm{min}$ & $T_\mathrm{b}^\mathrm{Gaus}$ & $W$ \\
		& (GHz) & & (K) & (K) & ($\umu$as) \\
		(1) & (2) & (3) & (4) & (5) & (6) \\\hline\hline
		2012-12-13 & 5 & 7.7 & $ 3.6\times10^{12} $ & $ 6.9\times10^{12} $ & 111.8 \\
		2012-12-15 & 5 & 14.9 & $ 3.8\times10^{12} $ & $ 1.9\times10^{13} $  & 71.4\\
		2012-12-15 & 22 & 14.8 & $ 1.6\times10^{13} $ & $ 3.8\times10^{13} $  & 13.3\\
		2012-12-16 & 5 & 18.7 & $ 8.4\times10^{12} $ & $ 3.2\times10^{13} $  & 54.4\\
		2013-08-27 & 5 & 2.2 & $ 6.9\times10^{11} $ & $1.0\times10^{12}$  & 376.2\\
		2013-08-27 & 1.7 & 2.8 & $ 2.3\times10^{12} $ & $ 2.4\times10^{12} $ & 457.3\\
		2015-01-15 & 1.7 & 19.8 & $ 2.9\times10^{12} $ & $ 2.0\times10^{13} $ & 164.4\\
		2015-12-16 & 5 & 8.7 & $ 1.1\times10^{13} $ & $ 1.8\times10^{13} $  & 95.7\\
		2015-12-17 & 5 & 16.4 & $ 1.4\times10^{13} $ & $ 5.9\times10^{13} $  & 63.1\\
		2015-12-29 & 5 & 25.5 & $ 1.4\times10^{13} $ & $ 9.2\times10^{13} $  & 43.3\\
		2016-01-09 & 5 & 14.5 & $ 1.8\times10^{13} $ & $ 4.1\times10^{13} $  & 62.0\\
		2016-01-29 & 5 & 8.9 & $ 6.4\times10^{12} $ & $ 1.8\times10^{13} $  & 104.6\\\hline
	\end{tabular}

\end{table}

The highest brightness temperature estimate comes from the fringe detection on the baseline 26 Earth diameters at 5\,GHz. The longest baseline in units of wavelengths is 14 G$\lambda$ achieved at 22\,GHz. The average apparent brightness temperature measured by \textit{RadioAstron} is about an order of magnitude higher than that measured by VLBA (Figure~\ref{fig:Tb_core_flux}). The estimates $T_\mathrm{b}^\mathrm{Gaus}$ and even conservative lower limit values $T_\mathrm{b}^\mathrm{min}$ do challenge the inverse Compton limit even after correction for boosting by the high value of $\delta\approx20$. 

\begin{figure}
	\centering
	\includegraphics[width=\columnwidth,trim=0cm 1cm 0cm 0cm]{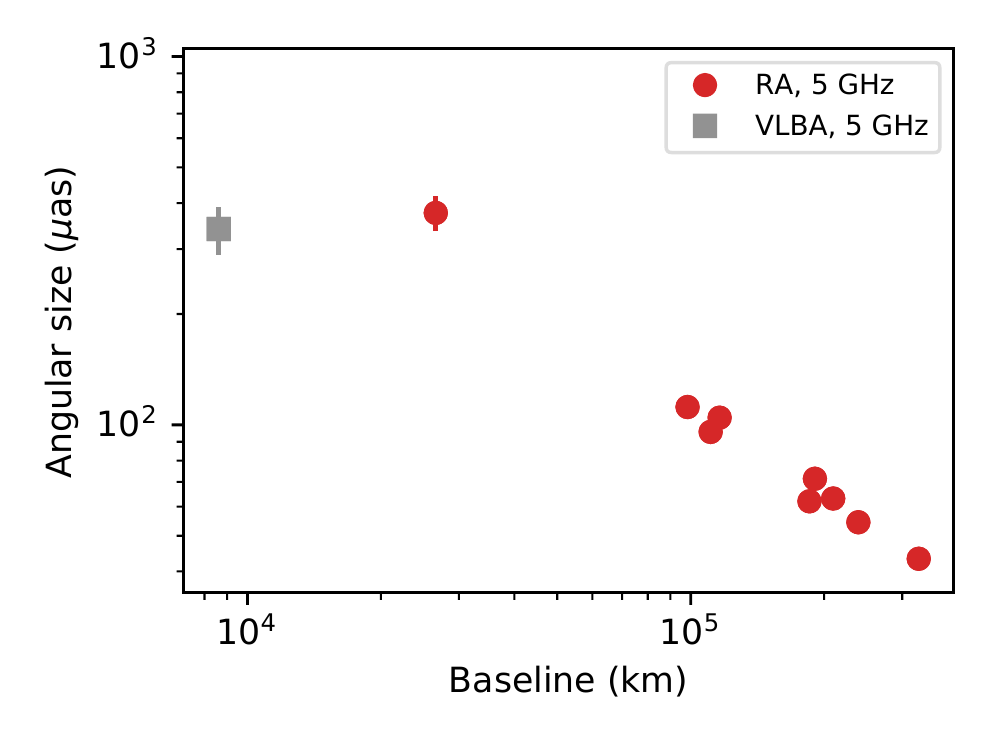}
	\caption{The source angular size estimated by fitting \textit{RadioAstron} visibility amplitudes at 5\,GHz (circles, Table~\ref{tab:ra}). The square marks the size of the core at 5\,GHz from ground-based VLBA measurements (Table~\ref{tab:difmap_models}).}
	\label{fig:size_ra}
\end{figure}

\begin{figure}
	\centering
	\includegraphics[width=\columnwidth,trim=0cm 1cm 0cm 0cm]{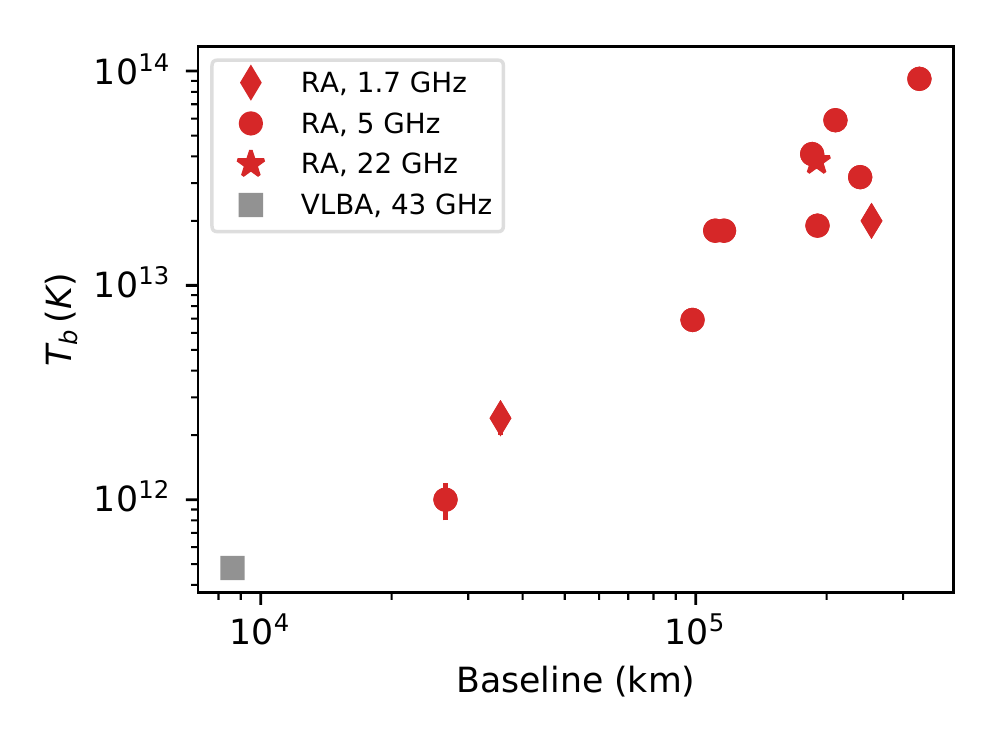}
	\caption{The apparent brightness temperature $T_\mathrm{b}^\mathrm{Gaus}$ measured by \textit{RadioAstron} vs.\ the baseline (Table~\ref{tab:ra}). Diamonds, circles, and the star represent $T_\mathrm{b}$ measured at 1.7, 5, and, 22~GHz, respectively. The square marks median $T_\mathrm{b}$ measured by 43\,GHz VLBA.}
	\label{fig:tb_ra}
\end{figure}

The size estimates $W$ at 5\,GHz are shown in Figure~\ref{fig:size_ra}. The core size obtained from modeling of the 5\,GHz VLBA data is also shown with the gray square. There is a plateau of the apparent source size on baselines less than about two Earth diameters, suggesting the presence of two scales in the inner compact source structure. The first one is the ``core''. It is resolved by ground-based VLBI and has intrinsic brightness temperature close to the equipartition value (see the previous Section), which can be considered as an average value over the whole core region. The second is ultra-compact, less than about $10\,\umu$as or 0.1\,pc, which remains unresolved even on the longest ground-space projected spacings. The brightness temperature, $T_\mathrm{b}^\mathrm{Gaus}$, measurements by \textit{RadioAstron} and VLBA are plotted against projected baseline in Figure~\ref{fig:tb_ra}. These estimates increase with projected baseline (see also Figure~\ref{fig:size_ra} for the corresponding angular size dependence) which is a clear indication of a lower limit behaviour. We conclude that the angular size values from Figure~\ref{fig:size_ra} and the brightness temperature values from Figure~\ref{fig:tb_ra} should be considered as upper and lower limits, correspondingly.

The high brightness temperatures measured by \textit{RadioAstron} 
in 0235+164 can be associated with an extremely compact ($<10\,\umu$as) feature in the jet of the source. It might be related to shocks or plasma blobs crossing the photosphere, e.g., through a tiny spine.
One can see that generally the measured brightness temperatures are higher when the source is in a flaring state (Figure~\ref{fig:Tb_core_flux}). However, the high values appear during the quiescent states as well. Therefore, the bright compact features occur in the jet without a prominent flare and their relation to the parsec-scale VLBI components (which are usually associated with the outbursts) is not clear.

\cite{2006ApJS..165..439R} have performed a comprehensive analysis of interstellar scintillation (ISS) in 0235+164. They indirectly estimated the size and flux density of an ultra-compact jet feature for 1992--1993 epochs and reported similar values. The authors found the increase of ISS amplitude during two major outbursts in the source (C and D in our notation) and suggest that the compact features might be associated with the new-born jet components. They travel down the jet and expand. The ISS of such a component has enough amplitude to produce the variability brightness temperatures up to $\sim 10^{14}$\,K at frequencies 2--8 GHz~\citep{2006ApJS..165..439R}.  

Interstellar scattering sub-structure \citep[e.g.,][]{2016ApJ...820L..10J} of these features, in principle, might lead to the observed values of brightness temperature at cm-wavelengths. However, it does not seem to play significant role in the observed source structure at 22\,GHz \citep[see for details][]{2015ApJ...805..180J}. Therefore, the values of $T_\mathrm{b}>10^{13}$\,K should have an intrinsic nature. We note that simultaneous measurements of $T_\mathrm{b}^\mathrm{Gaus}$ at 5\,GHz and 22\,GHz on 2012-12-15 (Table~\ref{tab:ra}) result in similar brightness temperature values providing an additional argument in favour of intrinsic scenario even at 5\,GHz. 

We find some clues to the presence of very fast bulk flow speed in the source (see section~\ref{sec:coreshift}). 
Possibly, there are regions with velocity gradients, like a spine-sheath structure of M87 \citep[e.g.,][]{2016A&A...595A..54M}. The higher flow speed in the spine could yield the higher Doppler boosting. Moreover, such a spine would be a supply channel of the high energy electrons, which can partially compensate the inverse Compton cooling~\citep[see e.g.,][]{1994ApJ...426...51R}. Then the high brightness level might persist in the source for a long time.
\cite{2016ApJ...820L...9K} discuss other possible explanations of the ultra-high apparent brightness temperatures, e.g., the synchrotron emission from relativistic protons, mono-energetic spectrum of particles, etc. However, they have significant difficulties in explaining the observational data.


\section{Jet structure}
\label{sec:jet}

\subsection{Image stacking}
\label{sec:stack}

\begin{figure}
	\centering{
	\includegraphics[width=\columnwidth,trim=0cm 1cm 0cm 0cm]{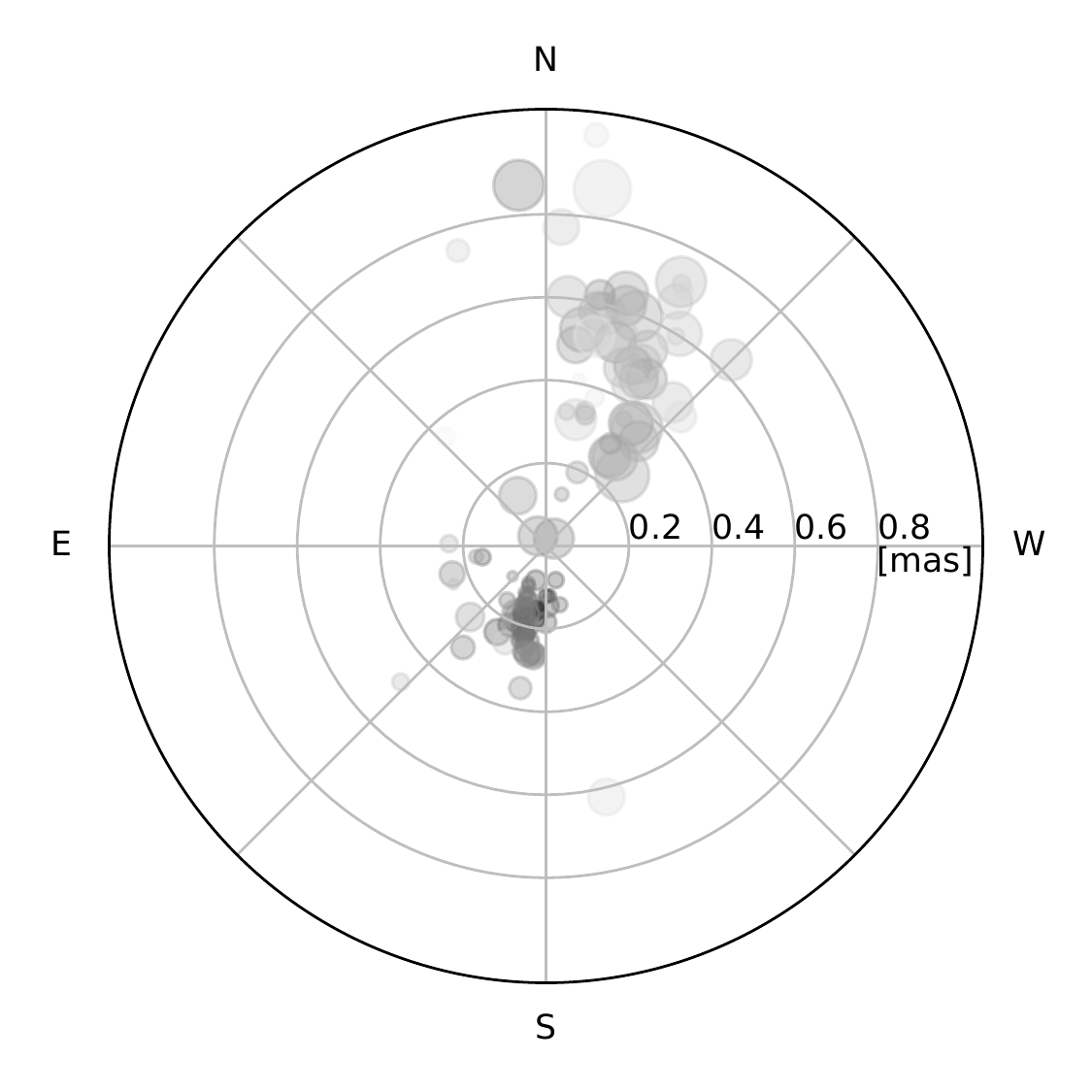}
	}
	\caption{The model components within 1 mas of the core, aligned relative to the position of the core (not plotted). The circles size and color intensity are proportional to the size and flux density of the components respectively.}
	\label{fig:structure_polar}
\end{figure}

\begin{figure}
	\includegraphics[width=0.54\columnwidth,trim=0cm 1cm 0cm 0cm]{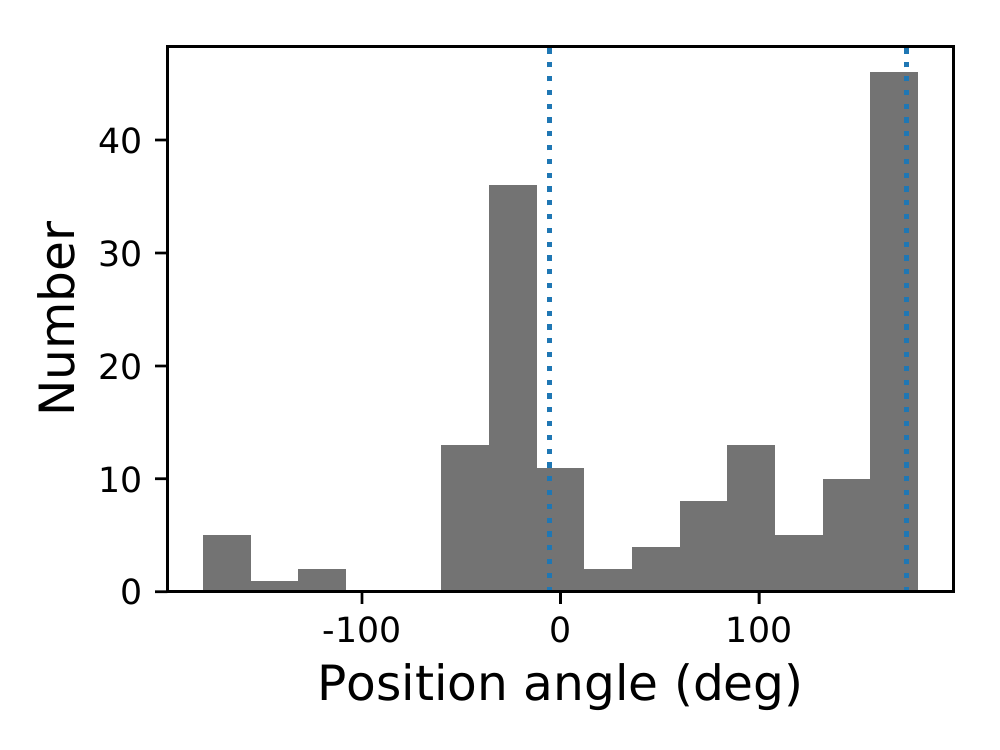}\includegraphics[width=0.54\columnwidth,trim=0cm 1cm 0cm 0cm]{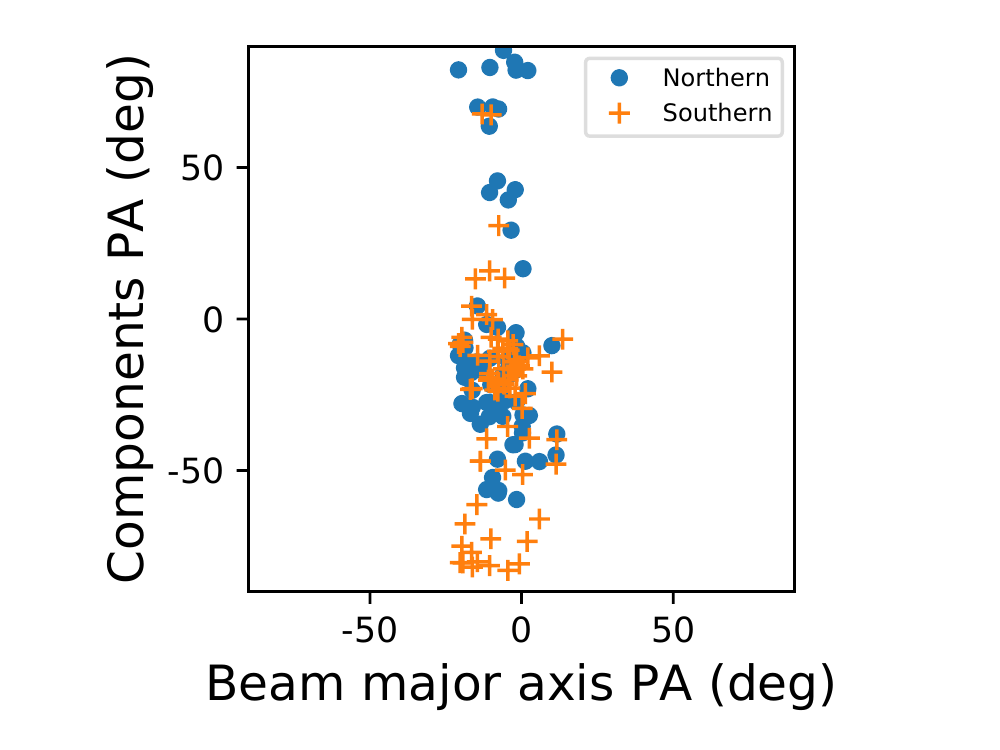}
	\caption{\textit{Left:} the distribution of the components PA (vertical lines show the median of the beam major axis orientation). \textit{Right:} the components PA reduced to $[-90^\circ, 90^\circ]$ vs.\ the PA of the beam major axis.}
	\label{fig:pa_distribution}
\end{figure}

\begin{figure}
	\includegraphics[width=\columnwidth,trim=0cm 1cm 0cm 0cm]{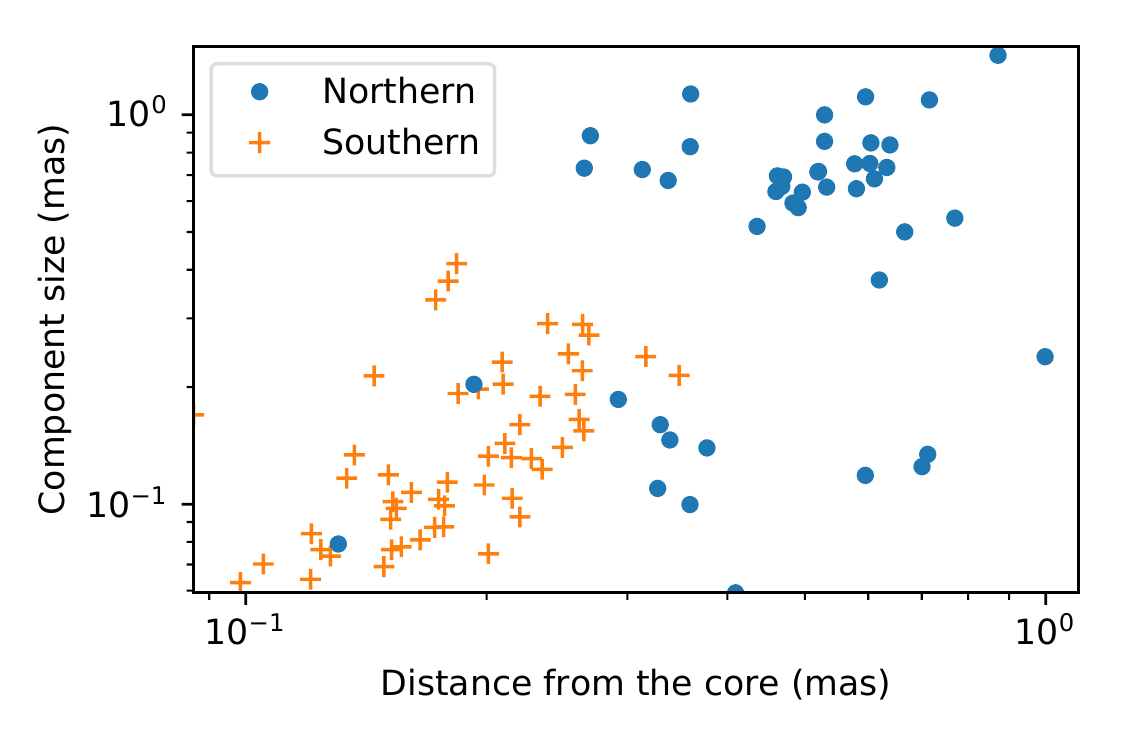}
	\caption{The size of components vs.\ their distance from the core.}
	\label{fig:sizes_dists}
\end{figure}

Due to the faint extended structure of 0235+164 it is difficult to determine jet geometrical characteristics from a single VLBI observation. Moreover, even the direction of the jet is not clear. In Figure~\ref{fig:structure_polar} all our models of the 100 43\,GHz VLBA observations are plotted with respect to the core position (the core is not shown). It is clearly seen that the components tend to concentrate almost symmetrically in two directions. We checked if the position angles (PA) of the components (median $-28^\circ$ and $162^\circ$) correlate with the major axis of the synthesized beam (median $-6^\circ$). They are found to be close but we did not find the correlation between them (Figure~\ref{fig:pa_distribution}). Therefore, the beam orientation does not seem to be responsible for the observed phenomena. Another support for the components PA indicating the jet PA comes from the fact that the size of the components increases with distance from the core both in southern and northern directions. Moreover, in the northern part the components are more distant and larger on average (see Figure~\ref{fig:sizes_dists}). We interpret this as a bend of the outflow from south to north, so that the younger (closer to the central engine) jet is observed southwards and the older (more extended and prominent) is seen northwards of the core.

\begin{figure}
	\includegraphics[width=\columnwidth,trim=0cm 0.5cm 0cm 0cm]{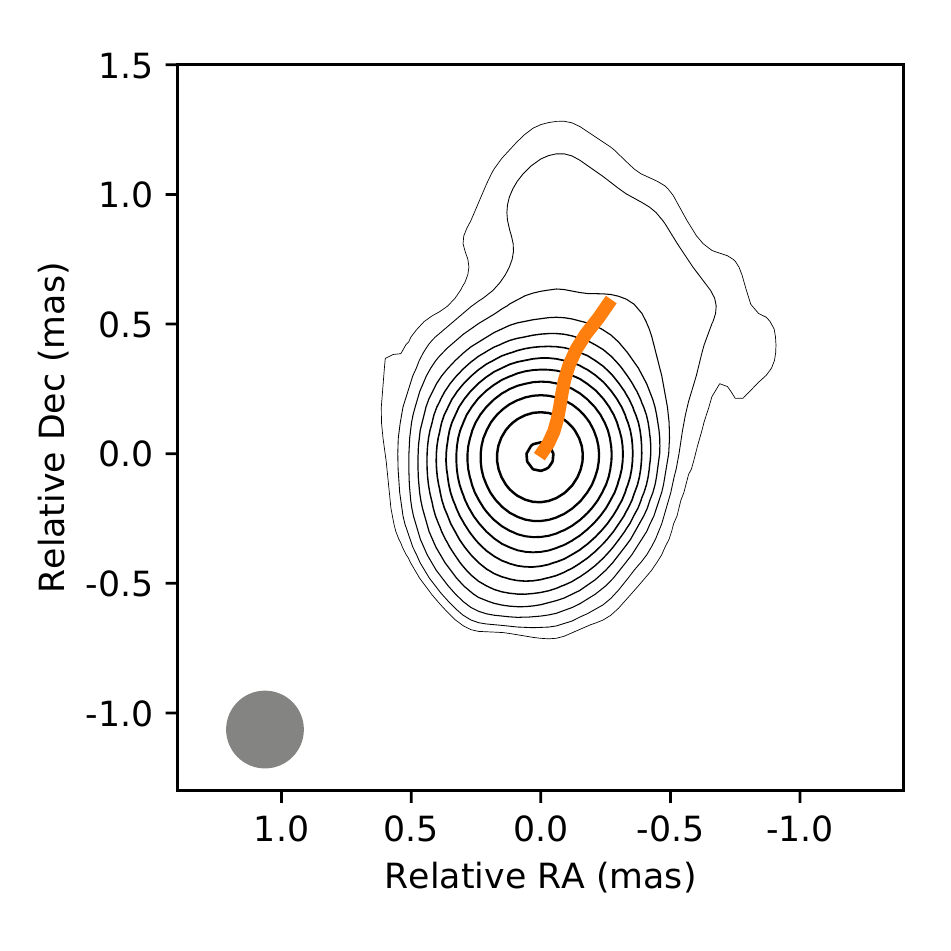}
	\caption{The stacked map of 0235+164 at 43\,GHz. Thick curve shows the ridge line. Grey circle in the lower left corner shows the beam at FWHM level. The contours of equal intensity are shown at x2 steps. The first contour is 0.6\,mJy/beam, the peak is 1.5\,Jy/beam, the RMS is 0.1\,mJy/beam.}
	\label{fig:ridgeline}
\end{figure}

\begin{figure}
	\includegraphics[width=\columnwidth,trim=0cm 1cm 0cm 0cm]{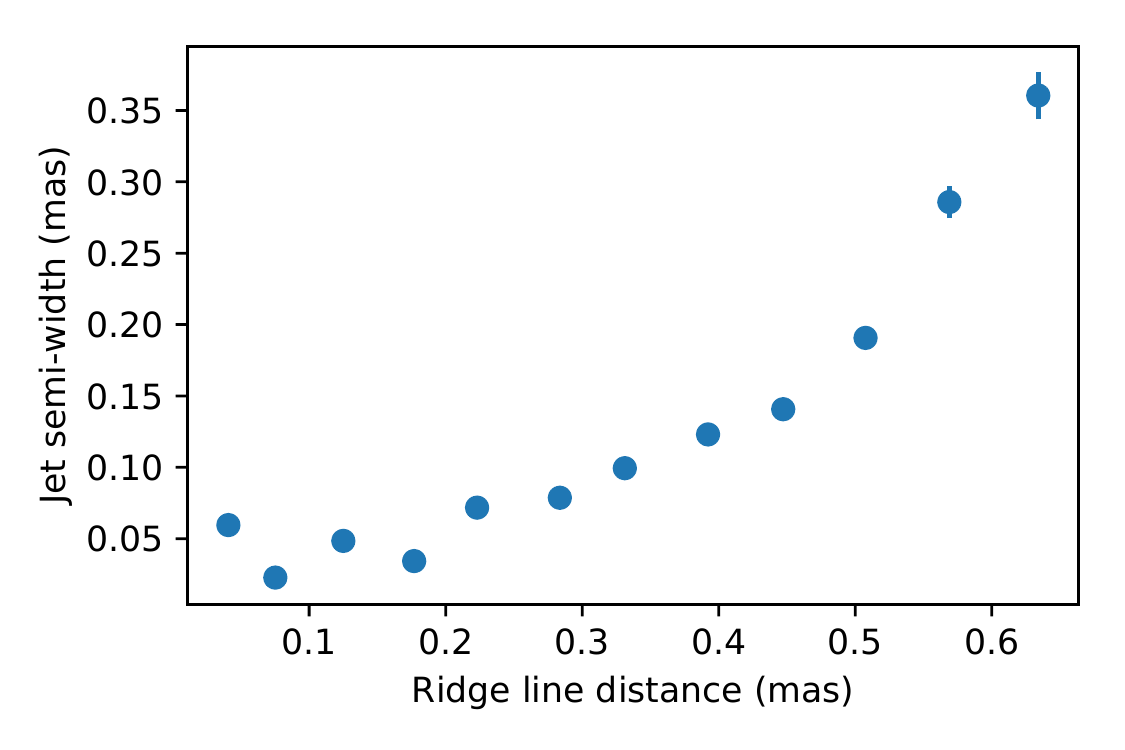}
	\caption{Jet half-width vs distance along the ridge line.}
	\label{fig:jet_width}
\end{figure}

We combine {a hundred of} 43\,GHz VLBA maps (obtained in independent experiments) aligned by the core position into a stacked image to increase dynamic range. For all the images we use the same circular beam {of 0.3\,mas at FWHM level}. This procedure allows us to detect the jet and to construct the ridgeline towards north-northwest of the core as shown in Figure~\ref{fig:ridgeline}. The jet to the south of the core is very compact and the corresponding ridgeline is too short to perform calculations. The intensity along transverse jet slices measured down the ridgeline is fitted with a Gaussian profile. The jet width $w$ is then found as a deconvolution of the obtained Gaussian with the beam.
The apparent opening angle $\varphi$ is estimated as $\varphi = 2\arctan (w/2l)$, where  $l$ is the distance along the ridgeline. The jet profile ($w/2$ vs.\ $l$) is shown in Figure~\ref{fig:jet_width}. 
{At the ridgeline distance longer than $0.7$\,mas the jet width estimate becomes uncertain due to low signal-to-noise ratio.}
Assuming conical geometry we find $\varphi_{43} = 60^\circ\pm 5^\circ$. It is twice as large as the value obtained by \cite{2017MNRAS.468.4992P} for stacked 15\,GHz image ($\varphi_{15} = 30^\circ$ on scales $\gtsim 1.5$\,mas). At 43\,GHz we observe the jet much closer to its apex, than at 15 GHz due to synchrotron self-absorption. Hence, we come to the conclusion that the flow is collimated 
within $\sim1.5$ mas of the apex. At lower frequency (1.4\,GHz, tens mas) the VLBA stacked image shows that the jet changes its direction to north-northeast \citep{2017MNRAS.468.4992P} keeping the opening angle $\sim 30^\circ$.

We note, however, that the above jet morphology is not the only one
consistent with the data presented. The absence of jet morphology
clearly visible in the image plane is consistent with a generic model
of a blazar with the direction of the jet outflow very close to the
direction to the observer. If such a geometry is considered in
combination with the assumption that what is perceived as the core is
actually a blend of a “true core” and a bright new variable (e.g.
quickly fading) component, then the centroid of that “assumed core” would
shift and produce apparent ``wobbling'' of the jet in opposite directions.
Further ingredients contributing into this morphology might
include a helical jet, e.g. as in the well-studied case of the blazar 1156$+$295 (\citealt{2004A&A...417..887H}, \citealt{2011A&A...529A.113Z}). In particular, the jet’s conical helicity with a varying pitch might explain the distribution of bright components by position angle relative to the core shown in Figure~\ref{fig:structure_polar}. Within the framework of this model, the jet components appearing on the opposite sides from the core would be at different spirals of the helical trajectory.

\subsection{Gc component}
\label{sec:component}

The G-flare (see Figure~\ref{fig:lightcurves}) is accompanied by a bright feature (Gc hereafter) located to the south-southeast from the core and traveling with apparent superluminal velocity (see below). It has been also studied by \cite{2011ApJ...735L..10A}, who interpreted it as a moving transverse shock. 

\begin{figure}
	\includegraphics[width=\columnwidth,trim=0cm 0.5cm 0cm 0cm]{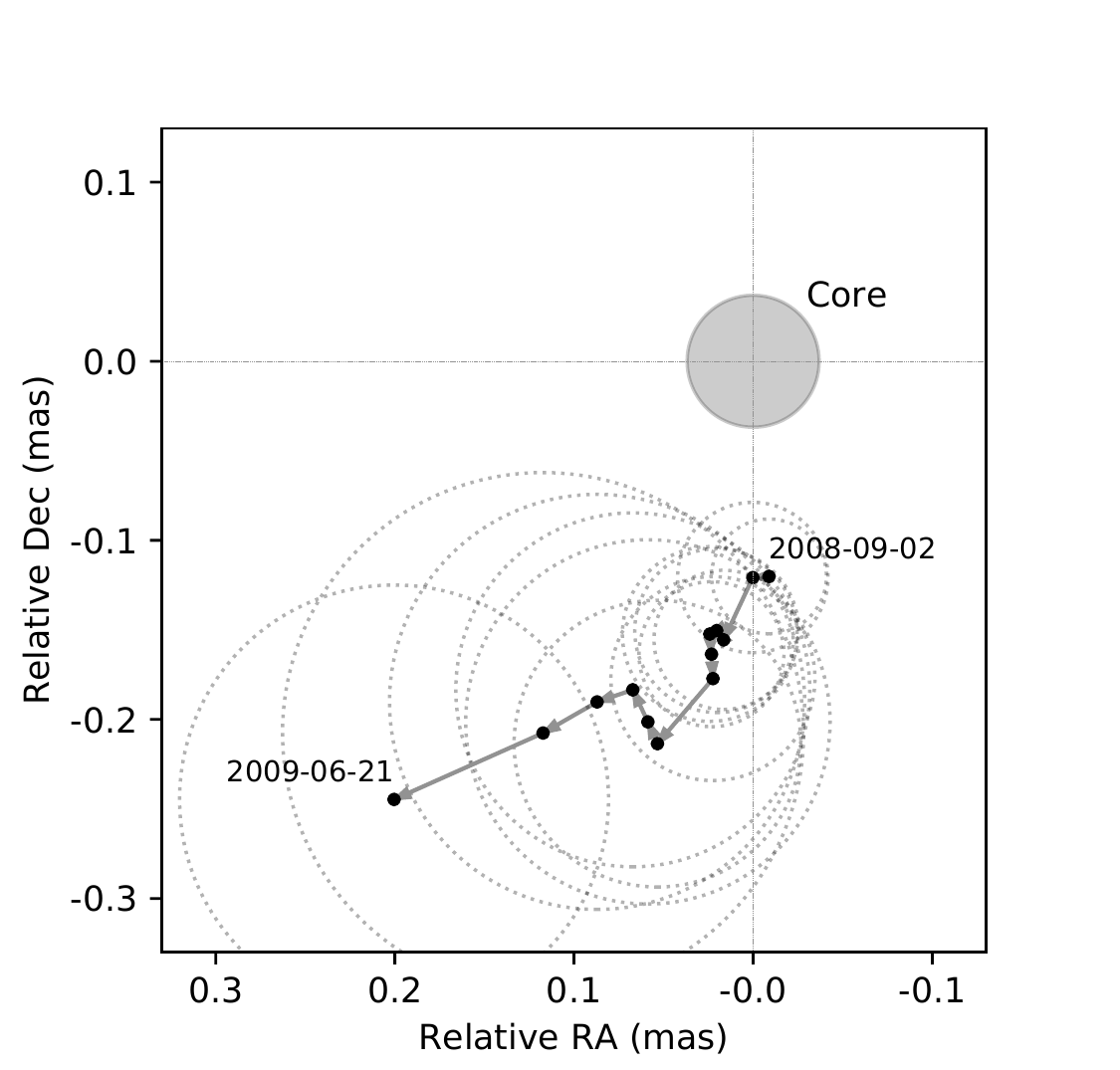}
	\caption{Trajectory of the Gc component during Sep 2008 -- Jun 2009 at 43 GHz. Dots, arrows and open circles show the position, path and size of the Gc. The core is at (0,0) and its median major axis shown with gray circle.} 
	\label{fig:Gc_radec}
\end{figure}

\begin{figure}
	\includegraphics[width=\columnwidth,trim=0cm 1cm 0cm 0cm]{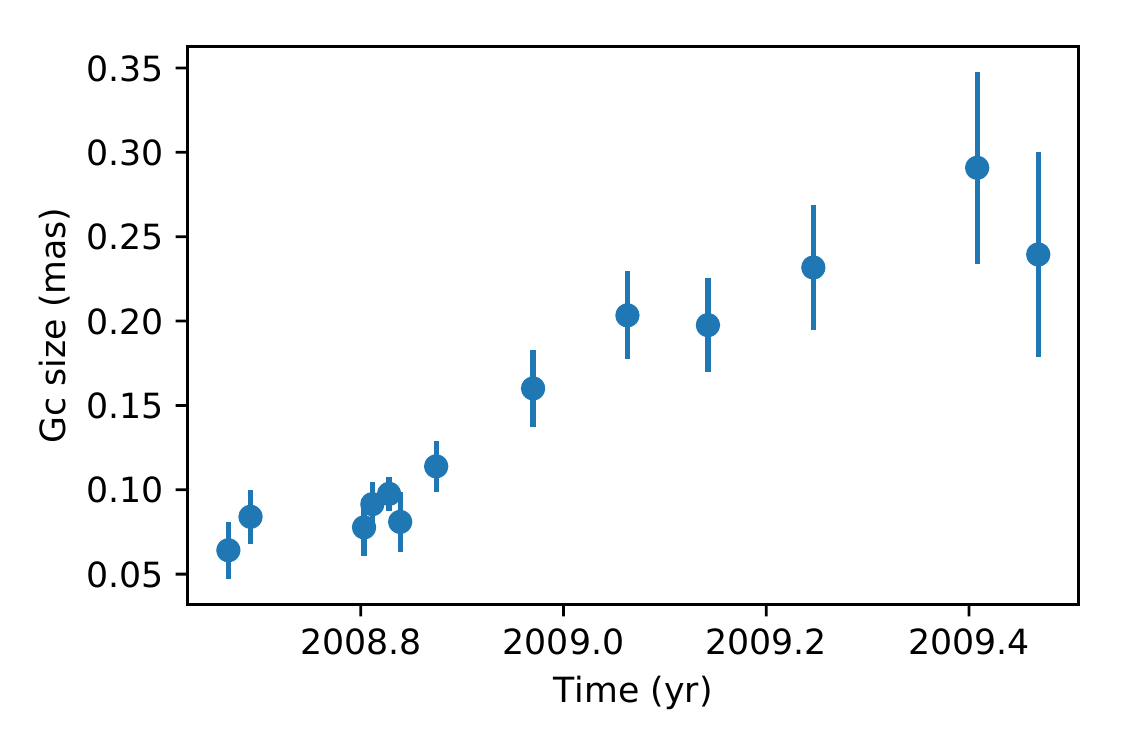}
	\caption{The size of the component Gc vs.\ time.}
	\label{fig:Gc_size_time}
\end{figure}

\begin{figure}
	\includegraphics[width=0.52\columnwidth,trim=0cm 1cm 0cm 0cm]{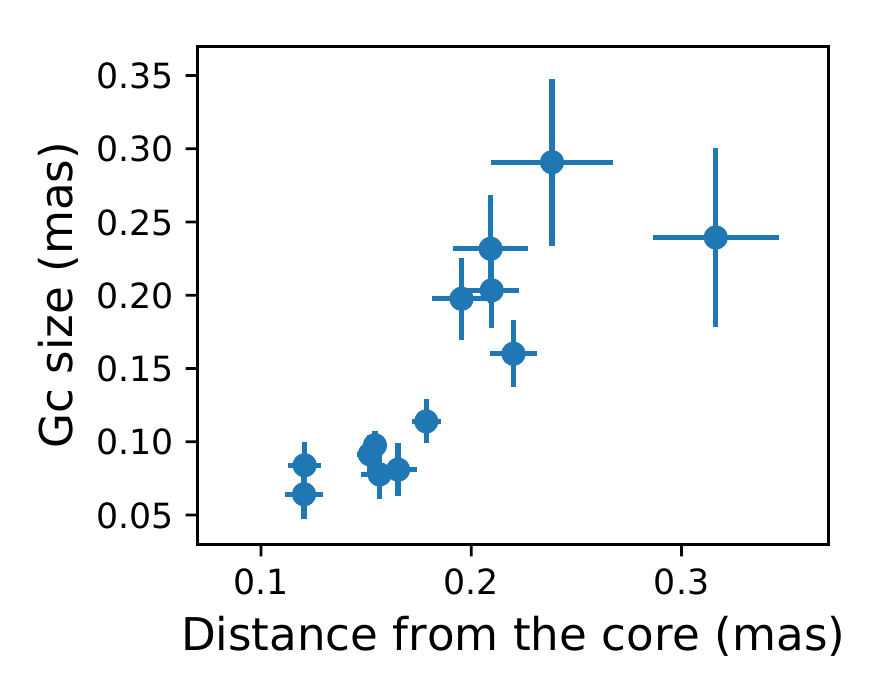}~
	\includegraphics[width=0.52\columnwidth,trim=0cm 1cm 0cm 0cm]{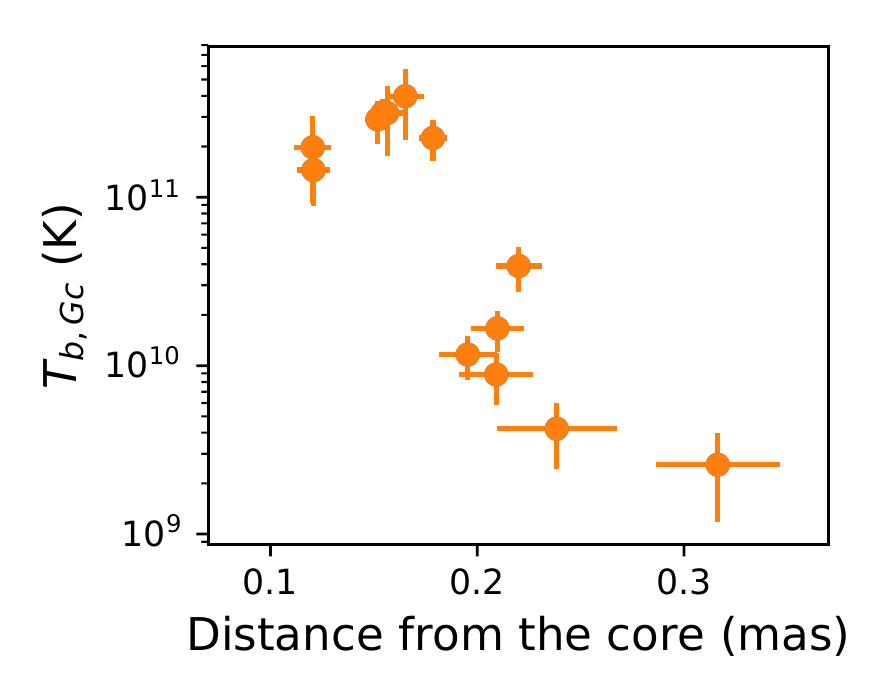}
	\caption{The size and apparent brightness temperature of Gc vs.\ its separation from the core.}
	\label{fig:Gc_size_tb}
\end{figure}

In Figure~\ref{fig:Gc_radec} the trajectory of Gc is shown relative to the core position. The mean proper motion of Gc relative to the core is $\umu = 0.20\pm 0.03$\,mas/yr, which corresponds to an apparent speed $\beta_{app} = 10\pm1.6\,c$. 
The position and speed of Gc is consistent with its ejection in Jan 2008, at the beginning of the core (and total flux) flare.

Using equations (\ref{eq:Tb}-\ref{eq:Tb_sys}) we estimate the brightness temperature and Doppler factor of Gc. During the flare peak $T_{b, Gc} \approx 4\times10^{11}$\,K (Figure~\ref{fig:Gc_size_tb}). Taking the variability time-scale at Gc-flare peak $t_{var, Gc} = 0.1$\,yr (estimated in the same way as for the core at flare decay in Nov 2008) and corresponding $T_{var, Gc} = 2.4\times10^{14}$\,K we obtain $T_{b, int, Gc} \approx 10^{10}$\,K and $\delta_{Gc} \approx 24$. 

The Lorentz factor and the viewing angle can be estimated using the following equations:

\begin{equation}
\Gamma = \frac{\beta_{app}^2 + \delta^2 + 1}{2\delta},\;\;\;\;\;
\theta = \arctan \frac{2\beta_{app}}{\beta_{app}^2 + \delta^2 - 1},
\end{equation}
which give $\Gamma \approx 14$ and $\theta \approx 1.7^\circ$. 

The slope of Gc size vs.\ distance from the core is $1.2\pm 0.4$ implying the conical angle $\varphi_{Gc} \approx 62^\circ$ is consistent with the opening angle $\varphi$ obtained from the stacked image in the northern jet. 
Therefore, the Gc component occupies the whole jet cross-section, supporting the bent-jet scenario. The intrinsic opening angle of the jet is then $\varphi_{int} = \varphi_{Gc} \sin\theta \approx 2^\circ$. 

If the viewing angle of the inner jet (in the core region) is the same as of Gc, then the Lorentz factor and apparent speed in that region are also kept as $\delta_{core}\approx\delta_{Gc}$. 

We come to the conclusion that the component Gc is the true physical part of the jet of 0235+164, rather than an artefact of calibrations. This is strongly supported by the evolution of its size and brightness temperature with time and distance from the core (Figures~\ref{fig:Gc_size_time},\ref{fig:Gc_size_tb}) as well as by its birth date. Moreover, since the component fits the width of the outflow, its trajectory depicts the form of the jet.

\citet{2017ApJ...846...98J} recently published the \texttt{Difmap} models of 0235+164 for 2007--2013 period based on the same data at 43\,GHz. The authors reported three components B1, B2, and B3 in the jet with estimated birth epochs 2007.4, 2008.3 and 2008.8 respectively. All of the features were found to have different velocities in different directions. The key difference between the aforementioned models and that obtained in this work is that we used elliptical Gaussian core based on the cross-validation. If the core has an elliptical form and circular model is used, one would typically obtain more than one very close and compact model-fit components. Some of them usually remain unresolved and have low flux density. These features needlessly complicate a model. We performed a brief comparison of the core flux density and size ($(W_{maj}*W_{min})^{1/2}$) derived in this work and that by \citet{2017ApJ...846...98J} and found good agreement of these parameters over all the period 2007--2013. We also used cross-validation to compare the models for 2008-08-15 observations epoch, where the core parameters mostly differ. The resulting cross-validation scores for the models are consistent within errors. At the same time model with elliptical core is simpler (two components) and hence is more preferable. We also note, that the same data were studied previously by \citet{2011ApJ...735L..10A}, who reported the single component in the jet of 0235+164 during 2008--2009.

\section{Core shift and physical parameters}
\label{sec:coreshift}

Due to the extreme compactness of the source the core shift can not be measured using optically thin jet {features}. However, it can be estimated indirectly based on the parameters derived above. The apparent core shift $\Delta r$ and change of its transverse size $\Delta W_\mathrm{min}$ as a function of frequency in a conical jet are related through the apparent opening angle $\varphi$ as $\Delta r = \Delta W_\mathrm{min} / 2\tan \varphi/2$. Here we assume that major and minor core axes are aligned with axial and transverse directions of the inner jet, which is typical for blazars~\citep{2005AJ....130.2473K}.

Then it is easy to estimate the expected core shift for a given frequency pair. 
For example, the shift between the core at 15\,GHz and 8\,GHz is $\Delta r_{15-8} \approx 0.03$\,mas, which is several times lower than the median value (0.128 mas) obtained by \cite{2012A&A...545A.113P} at these frequencies. The smaller core shift value is expected for a source having a small viewing angle. The core shift between 15\,GHz and 5\,GHz should be $\Delta r_{15-5} \approx 0.1$\,mas. 

Consider the standard scalings for the particle density $N=N_1(r/r_1)^{-n}$ and the magnetic field strength $B = B_1(r/r_1)^{-m}$, where index $1$ refers to values at 1~pc from the central engine. In a freely expanding jet (without ambient-medium pressure) the power-law indices $m$, $n$ can be linked through the index $k$ (the conical geometry implies that the core shift index $k$ equals to $k_w$ measured in Section~\ref{sec:core}) and the optically thin spectral index $\alpha$~\citep{1981ApJ...243..700K}\footnote{Note, that throughout this paper $k$ is defined as K\"{o}nigl's $k_r^{-1}$.}:

\begin{equation}\label{eq:k}
k=\frac{5-2\alpha}{(3-2\alpha)m+2n-2}
\end{equation}

Further we discuss the equipartition regime, when the energy density of relativistic electrons roughly equals to that of the magnetic field. We also consider another scenario, where the conical jet undergoes adiabatic losses. 

\subsection{Equipartition case}

Estimation of physical parameters can be performed taking into account the fact that the intrinsic brightness temperature of the core is close to the equipartition value $T_\mathrm{b,eq}\sim10^{10.5}$\,K. 
The equipartition between particles and magnetic field energy density implies that $n=2m$, and from (\ref{eq:k}) for $k=0.8\pm0.1$, $\alpha=-0.5$ we obtain $m = 1.2\pm0.1$ and $n = 2.4\pm0.2$. Note, that the dependence of the parameters on $\alpha$ is weak, and using $\alpha=-0.75$ changes the above results insignificantly.

The measure of core offset for a given frequency pair $\nu_1, \nu_2$ can be introduced following \citet{1998A&A...330...79L}:

\begin{equation}\label{eq:Omega}
\Omega=4.85\times10^{-9}\frac{\Delta
	r_\mathrm{mas}D_\ell}{(1+z)^2}\left(\frac{\nu_1^{k}\nu_2^{k}}{\nu_2^{k}-\nu_1^{k}}\right),
\end{equation}
where $D_\ell$ is the luminosity distance in pc, $\Delta r_\mathrm{mas}$ -- the core shift in mas. Then magnetic field strength at 1\,pc:

\begin{equation}\label{eq:B1}
B_{1}\approx
0.025\, \mathrm{G}\left(\frac{\Omega^{3/k}(1+z)^2}{\delta^2\phi\sin^{3/k-1}\theta}\right)^{1/4},
\end{equation}
where $\theta$~is the jet viewing angle, $\phi$ is the jet half
opening angle and $\delta$ is the Doppler factor. Substitution of the parameters gives $B_1 \approx 1.3$\,G. The de-projected distance from the central engine to the core at a frequency $\nu$ is then:

\begin{equation}\label{eq:rcore}
r_{\rm core} (\nu)
= \frac{\Omega}{\sin \theta} \nu^{-k} \,\,\mathrm{[pc]}
\end{equation}

E.g., the core at 43\,GHz is located at $r_{\rm core, 43} \approx 10.5$\,pc, while the core at 2.3\,GHz is more than 100 pc downstream from the jet apex. The magnetic field in the 43\,GHz core is $B_\mathrm{43,core}\approx0.1$\,G, which is typical for blazars (e.g., \citealt{2011A&A...532A..38S, 2012A&A...545A.113P, 2014MNRAS.437.3396K, 2017MNRAS.468.4478L}).

\subsection{Adiabatic case}

The inverted spectrum of the core at lower frequencies (Figure~\ref{fig:corespec}) may also be related to the adiabatic losses, which might dominate in the jet on parsec scales~\citep{1980ApJ...235..386M}. 
Let $m=1$, as expected for the transverse magnetic field~\citep{1974MNRAS.169..395B}, and $n=2/3(2-s)$, where $s$ -- is the slope of the electron energy distribution's power law. The optically thin spectral index is $\alpha = (s + 1)/2$. Then the solution of Eq.~(\ref{eq:k}) gives $s=-2.15$. This value is very close to that predicted by models of electron acceleration by relativistic shocks in AGN jets~\citep{2000ApJ...542..235K}. Note, that the dependence of $s$ on the observed index $k$ is very strong, e.g., changing $k$ from 0.7 to 0.9 changes $s$ from -3.4 to -1.5. If we assume $m=2$ instead (the longitudinal magnetic field), we obtain unrealistic slope $s(k=0.8) = -0.16$. 

Therefore, both equipartition and adiabatic cases suggest $m$ close to 1, and reject $m=2$, implying the dominance of the transverse component of the magnetic field in the parsec-scale jet. 

\subsection{Jet speed}
\label{sec:jet_speed}

In Section~\ref{sec:component} we estimate the apparent speed of the superluminal component. The obtained value is typical among the velocities derived from kinematics of the moving features in the jets of luminous blazars~\citep[see e.g., Fig.\,10 by][]{2007Ap&SS.311..231K}. There is yet another method to estimate apparent speed of a jet at its innermost parts (in the core region).

It is natural to assume that the peak of a flare at a given frequency occurs when a disturbance travelling downstream the jet crosses the core at this frequency. Then the apparent core shift can be expressed through the time delay as $\Delta r = \umu\Delta T$, where $\umu$ is the proper motion of the bulk flow in the core region.
For $\Delta r_{15-5} = 0.1$\,mas and $\Delta T_\mathrm{G, 15-5} = 14$\,days we obtain $\umu=2.6$~mas/yr ($130\,c$ at the source redshift), which is an order of magnitude higher than the proper motion of the Gc jet component. 
Therefore, the bulk flow speed is much higher than the pattern speed in the source outflow. Similar result was obtained for the blazar 3C\,454.3 by~\citet{2014MNRAS.437.3396K}. The observed extreme brightness temperatures of the innermost jet regions may be attributed to the ultra-high bulk flow speed. We note, however, that the core shift of 0235+164 is estimated indirectly, and might be a subject of concern. {In any case, this question must be studied further by direct core shift and time delays measurements of more blazars.}

\section{Summary}
\label{sec:conclusions}

We study the frequency dependence of peak-to-peak time delays of the outbursts in the blazar AO\,0235+164. In parallel, we analyze the frequency dependence of the apparent size of the radio core in the source. Our results support the dominance of synchrotron self-absorption in the jet at cm-wavelengths. 
We implement a self-consistent method to estimate Doppler factor and intrinsic brightness temperature of the core using the size and variability scale of the same region. The resulting value of $T_\mathrm{b,int}$ is close to that expected in the equipartition regime. 

The brightness temperature measured using ground-based VLBI increases by a factor of about 2 during the flares, but still is much lower than that obtained with the ground-space radio interferometer \textit{RadioAstron}, where visibility amplitudes of 50\,mJy to 100\,mJy are detected on baselines up to 14~G$\lambda$. We find evidence for the presence of ultra-compact, less than 10\,$\umu$as or 0.1\,pc, features in the source. 
They might demonstrate fast flux density variations at cm-wavelengths via the ISS mechanism, in good agreement with the observed intra-day variability of 0235+164. 
The recently discovered scattering sub-structure is not expected to appear at least at 22\,GHz, therefore, the extreme brightness of the source up to $10^{14}$\,K might have intrinsic nature, and is possibly related to regions of ultra-high bulk flow speed in the inner jet. The signature of high speed in the jet base also comes from the comparison of the derived core shift and the corresponding peak-to-peak time delay at frequencies 5 and 15 GHz for the flare in 2008. 

There is a superluminal jet feature in the source seen on 43\,GHz maps during the flare.
The estimated birth epoch as well as the increase of its size with time and distance from the core support the conclusion that it is real jet component. The Doppler factor of the component estimated from its kinematic analysis is in good agreement with that found for the core. 
The two prevalent directions in the spread of the components at 43\,GHz are probably caused by jet bending from south to north at about 0.3\,mas from the core. The brighter, more compact components are observed closer to the core in the southern jet during all the period studied, which provides strong support for this scenario. The estimates of the opening angle of the southern (from the component size) and the northern (from the stacked image) jet also suggest that its direction changes. Millimeter interferometric observations with higher resolution in the north-south direction would shed more light on this structure. 
Our estimates of the opening angle, compared to the previously reported, suggest the collimation of the outflow within 1.5\,mas of the central engine. Inside this region the intrinsic opening angle of the jet cone is close to the viewing angle, yielding the additional dispersion of position angles of the observed components. 
We estimate high, but not extreme values of the Lorentz factor $\Gamma = 14$ and the Doppler factor $\delta=21$, and viewing angle $\theta = 1.7^\circ$. 

Based on the derived jet geometry we estimate the expected core shift in the jet of 0235+164. There is evidence that the bulk plasma speed is an order of magnitude higher than the pattern speed in the jet. We consider the equipartition and adiabatic scenarios, which adequately describe the observational data. Both cases imply the dominance of a transverse magnetic field component. We also derive the gradients of magnetic field strength and electron density in the jet, as well as the distance from the jet apex to the core at each frequency.

\section*{Acknowledgments}

We thank Alan Roy, Alexander Pushkarev, Alan Marscher,  David Jauncey and the anonymous reviewer for useful comments which helped to improve the manuscript.
This research was supported by Russian Science Foundation (project 16-12-10481).
The \textit{RadioAstron} project is led by the Astro Space Center of the Lebedev Physical Institute of the Russian Academy of Sciences and the Lavochkin Scientific and Production Association under a contract with the Russian Federal Space Agency, in collaboration with partner organizations in Russia and other countries. Results of optical positioning measurements of the Spektr-R spacecraft by the global MASTER Robotic Net \citep{2010AdAst2010E..30L}, ISON collaboration, and Kourovka observatory were used for spacecraft orbit determination in addition to mission facilities.
This research is partly based 
on the Evpatoria RT-70 radio telescope (Ukraine) observations carried out by the Institute of Radio Astronomy of the National Academy of Sciences of Ukraine under a contract with the State Space Agency of Ukraine and by the National Space Facilities Control and Test Center with technical support by Astro Space Center of Lebedev Physical Institute, Russian Academy of Sciences.
This research is partly based 
on the observations with the 100-m telescope of the MPIfR (Max-Planck-Institute for Radio Astronomy) at Effelsberg.
This research is partly based 
on the observations with the Arecibo Observatory, which is operated by SRI International under a cooperative agreement with the National Science Foundation (AST-1100968), and in alliance with Ana G. Mendez-Universidad Metropolitana, and the Universities Space Research Association.
This research is partly based 
on the observations carried out using the 32-meter radio telescope operated by Torun Centre for Astronomy of Nicolaus Copernicus University in Torun (Poland) and supported by the Polish Ministry of Science and Higher Education SpUB grant.
This research is partly based 
on the observations with the Medicina and Noto telescopes operated by INAF - Istituto di Radioastronomia.
The Green Bank Observatory is a facility of the National Science Foundation operated under cooperative agreement by Associated Universities, Inc.
The Long Baseline Observatory is a facility of the National Science Foundation operated by Associated Universities, Inc. 
The European VLBI Network is a joint facility of independent European, African, Asian, and North American radio astronomy institutes. Scientific results from data presented in this publication are derived from the EVN project EK028.
The study has made use of data from the OVRO 40-m monitoring program \citep{2011ApJS..194...29R} which is supported in part by NASA grants NNX08AW31G, NNX11A043G, and NNX14AQ89G and NSF grants AST-0808050 and AST-1109911.
The paper makes use of 43\,GHz VLBA data from the VLBA-BU Blazar Monitoring Program (VLBA-BU-BLAZAR\footnote{\url{http://www.bu.edu/blazars/VLBAproject.html}}), funded by NASA through the Fermi Guest Investigator Program. The VLBA is an instrument of the Long Baseline Observatory.
UMRAO was supported in part by a series of grants from the NSF, most recently AST-0607523, and by NASA Fermi Guest Investigator grants NNX09AU16G, NNX10AP16G, NNX11AO13G, and NNX13AP18G.

\bibliographystyle{mnras}
\bibliography{kutkin_0235}

\label{lastpage}
\end{document}